\documentclass{article} 
\usepackage{iclr2021_conference,times}

\usepackage{amsmath,amsfonts,bm}

\def\eqref#1{equation~\ref{#1}}

\def\1{\bm{1}}

\DeclareMathAlphabet{\mathsfit}{\encodingdefault}{\sfdefault}{m}{sl}
\SetMathAlphabet{\mathsfit}{bold}{\encodingdefault}{\sfdefault}{bx}{n}

\usepackage{hyperref}
\usepackage{url}

\usepackage{caption,subcaption}
\usepackage{multirow}
\usepackage{float} 
\usepackage{hyperref}
\usepackage{pifont}
\usepackage{graphicx}
\usepackage{pgfplots}
\usetikzlibrary{spy}

\definecolor{palet_red}{RGB}{230, 57, 70}
\definecolor{palet_white}{RGB}{241, 250, 238}
\definecolor{palet_light_blue}{RGB}{168, 218, 220}
\definecolor{palet_medium_blue}{RGB}{69, 123, 157}
\definecolor{palet_dark_blue}{RGB}{29, 53, 87}
\definecolor{emerald}{rgb}{0.31, 0.78, 0.47}
\definecolor{nored}{RGB}{219,68,55}
\definecolor{yesgreen}{RGB}{13,140,79}

\newcommand{\imagepath}{./arxiv_figure}

\newcommand{\pastref}{\hat{\mathbf{x}}_{p}}
\newcommand{\futureref}{\hat{\mathbf{x}}_{f}}
\newcommand{\pastflow}{\mathbf{v}_{p}}
\newcommand{\futureflow}{\mathbf{v}_{f}}
\newcommand{\balph}{\boldsymbol{\alpha}}
\newcommand{\bbeta}{\boldsymbol{\beta}}
\newcommand{\pred}{\tilde{\mathbf{x}}_t}
\newcommand{\sysoutput}{\hat{\mathbf{x}}_t}

\title{Conditional Coding for Flexible Learned Video Compression}

\author{Th\'{e}o Ladune \& Pierrick Philippe \\
Orange, Rennes, France \\
\footnotesize{\texttt{firstname.lastname@orange.com}} \\
\And
Wassim Hamidouche, Lu Zhang \& Olivier D\'{e}forges \\
Univ. Rennes, INSA Rennes, CNRS, IETR UMR 6164 \\
\footnotesize{\texttt{firstname.lastname@insa-rennes.fr}} \\
}

\iclrfinalcopy 
\begin{document}

\maketitle

\begin{abstract}
This paper introduces a novel framework for end-to-end learned video coding.
Image compression is generalized through conditional coding to exploit information
from reference frames, allowing to process intra and inter frames with the same
coder. The system is trained through the minimization of a rate-distortion cost,
with no pre-training or proxy loss. Its flexibility is assessed under three
coding configurations (All Intra, Low-delay P and Random Access), where it is
shown to achieve performance competitive with the state-of-the-art video codec
HEVC. 
\end{abstract}

\section{Introduction}

In the last few years, ITU/MPEG video coding standards---HEVC
\citep{Sullivan:2012:OHE:2709080.2709221} and VVC \citep{VVC_Ref}---have been
challenged by learning-based codecs. The learned image coding framework
introduced by \citet{DBLP:conf/iclr/BalleLS17,DBLP:conf/iclr/BalleMSHJ18} eases
the design process and improves the performance by jointly optimizing all steps
(encoder, decoder, entropy coding) given a rate-distortion objective. The best
learned coding system \citep{cheng2020learned} exhibits performance on par with
the image coding configuration of VVC. In video coding, temporal redundancies
are removed through motion compensation. Motion information between
frames are transmitted and used to interpolate reference frames to obtain a
temporal prediction. Then, only the residue (prediction error) is sent, reducing
the rate. Frames coded using references are called \textit{inter} frames, while others
are called \textit{intra} frames.

Although most learning-based video coding systems follow the framework of
Ball\'{e} et al., the end-to-end character of the training is often overlooked.
The coders introduced by \citet{DBLP:conf/cvpr/LuO0ZCG19} or
\citep{DBLP:journals/corr/abs-1912-06348} rely on a dedicated pre-training to
achieve efficient motion compensation. Dedicated training requires proxy
metrics not necessary in line with the real rate-distortion objective, leading
to suboptimal systems. Due to the presence of both intra and inter frames,
learned video coding methods transmit two kinds of signal: image-domain signal
for intra frames and residual-domain for inter frames. Therefore, most works
\citep{Agustsson_2020_CVPR} adopt a \textit{two-coder} approach, with separate
coders for intra and inter frames, resulting in heavier and less factorizable
systems.

This paper addresses these shortcomings by introducing a novel framework for
end-to-end learned video coding, based on a single coder for both intra and
inter frames. Pursuing the work of \citet{LaduneMMSP20}, the coding scheme is
decomposed into two sub-networks: MOFNet and CodecNet. MOFNet conveys motion
information and a coding mode, which arbitrates between transmission with
CodecNet or copy of the temporal prediction. MOFNet and CodecNet use conditional
coding to leverage information from the previously coded frames while being
resilient to their absence. This allows to process intra and inter frames with
the same coder. The system is trained as a whole with no pre-training or
dedicated loss term for any of the components. It is shown that the system is
flexible enough to be competitive with HEVC under three coding configurations.

\vspace{-0.03cm}

\section{Proposed system}

\begin{figure}[ht]
    \centering
    \includegraphics[width=0.6\linewidth]{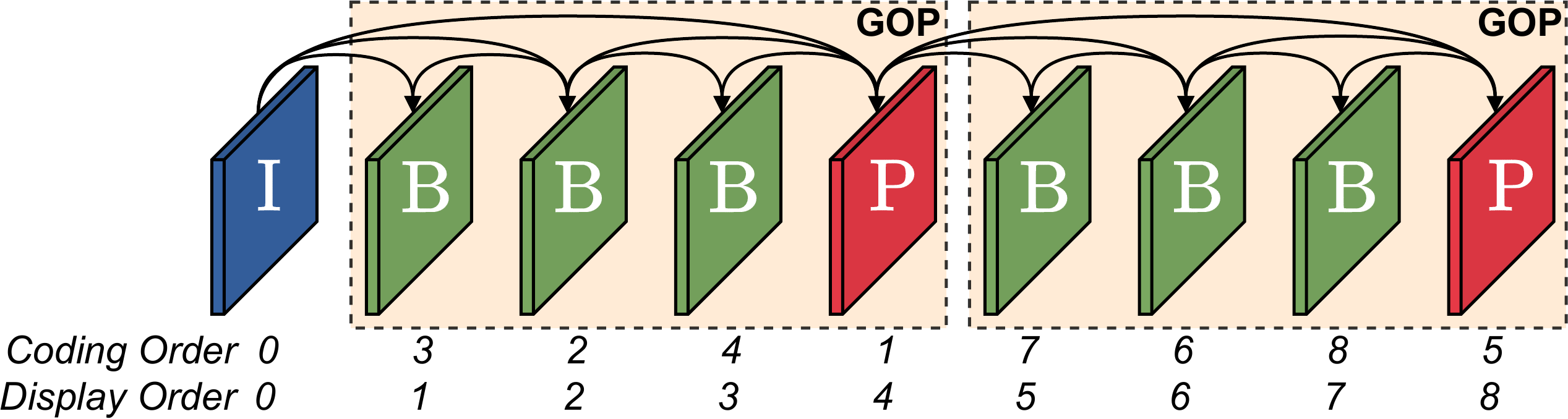}
    \caption{Random Access configuration, GOP size is set to 4 to have concise diagrams.}
    \label{fig:gopstruct}
\end{figure}


Let $\left\{\mathbf{x}_i, i \in \mathbb{N}\right\}$ be a video sequence, each
frame $\mathbf{x}_i$ being a vector of $C$ color channels\footnote{Videos are in
YUV 420. For convenience, a bilinear upsampling is used to obtain YUV 444 data.}
of height $H$ and width $W$. Video codecs usually process Groups Of Pictures
(GOP) of size $N$, with a regular frame organization. Inside a GOP, all frames
are inter-coded and rely on already sent frames called references: B-frames use
two references while P-frames use a single one. The first frame of the GOP
relies either on a preceding GOP or on an intra-frame (I-frame) denoted as
$\mathbf{x}_0$. This work primarily targets the \textit{Random Access}
configuration (Fig. \ref{fig:gopstruct}), because it features I, P and B-frames.
Here, we consider the rate-distortion trade-off, weighted by $\lambda$, of a
\textit{single} GOP plus an initial I-frame $\mathbf{x}_0$:
\begin{equation}
    \mathcal{L}_\lambda = \sum_{t=0}^{N} \mathrm{D}(\hat{\mathbf{x}}_t, \mathbf{x}_t) +
    \lambda \mathrm{R}(\hat{\mathbf{x}}_t), \text{ with }\mathrm{D} \text{ the MSE and } \mathrm{R} \text{ the rate.}
    \label{eq:loss}
\end{equation}

\subsection{B-frame Coding}

The proposed architecture processes the entire GOP (I, P and B-frames) using a
unique neural-based coder. B-frames coding is detailed here. Thanks to
conditional coding, I and P-frames are processed by simply bypassing some steps
of the B-frame coding process as explained in Section
\ref{subsec:conditionalcoding}.

Let $\mathbf{x}_t$ be the current B-frame and $(\pastref,\futureref)$ two
reference frames. Figure \ref{fig:overalldiagram} depicts the coding process of
$\mathbf{x}_t$. First, $(\mathbf{x}_t,\pastref,\futureref)$ are fed to MOFNet
which computes and conveys---at a rate  $R_m$---two optical flows $(\pastflow,
\futureflow)$, a pixel-wise prediction weighting $\bbeta$ and a pixel-wise
coding mode selection $\balph$. The optical flow $\pastflow$ (respectively
$\futureflow$) represents a 2D pixel-wise motion from $\mathbf{x}_t$ to
$\pastref$ (resp. $\futureref$). It is used to interpolate the reference through a
bilinear warping $w$. The pixel-wise weighting $\bbeta$ is applied to obtain the
bi-directional weighted prediction $\pred$:
\begin{equation}
    \pred = \bbeta \odot w(\pastref; \pastflow) + (1 - \bbeta) \odot w(\futureref; \futureflow),
    \left\{ \begin{array}{l}
        \odot \text{ is a pixel-wise multiplication,} \\
        \pastflow \text{ and } \futureflow \in \mathbb{R}^{2 \times H \times W},\ \bbeta  \in \left[0, 1\right]^{H \times W}
    \end{array}
    \right.
    \label{eq:pred}
\end{equation}
The coding mode selection $\balph \in \left[0, 1\right]^{H \times W}$ arbitrates
between transmission of $\mathbf{x}_t$ using CodecNet versus \textit{Skip mode},
a direct copy of $\pred$. CodecNet sends areas of $\mathbf{x}_t$ selected by
$\balph$, using information from $\pred$ to reduce its rate $R_c$. The total
rate required for $\mathbf{x}_t$ is $R = R_m + R_c$ and the decoded frame
$\sysoutput$ is the sum of both contributions: $\sysoutput = \underbrace{(1 -
\balph) \odot \pred}_{\text{Skip}} + \underbrace{c(\balph \odot \mathbf{x}_t,
\balph \odot \pred)}_{\text{CodecNet}}$.

\subsection{Conditional Coding}
\label{subsec:conditionalcoding}

\begin{figure}[ht]
    \centering
    \includegraphics[width=\linewidth]{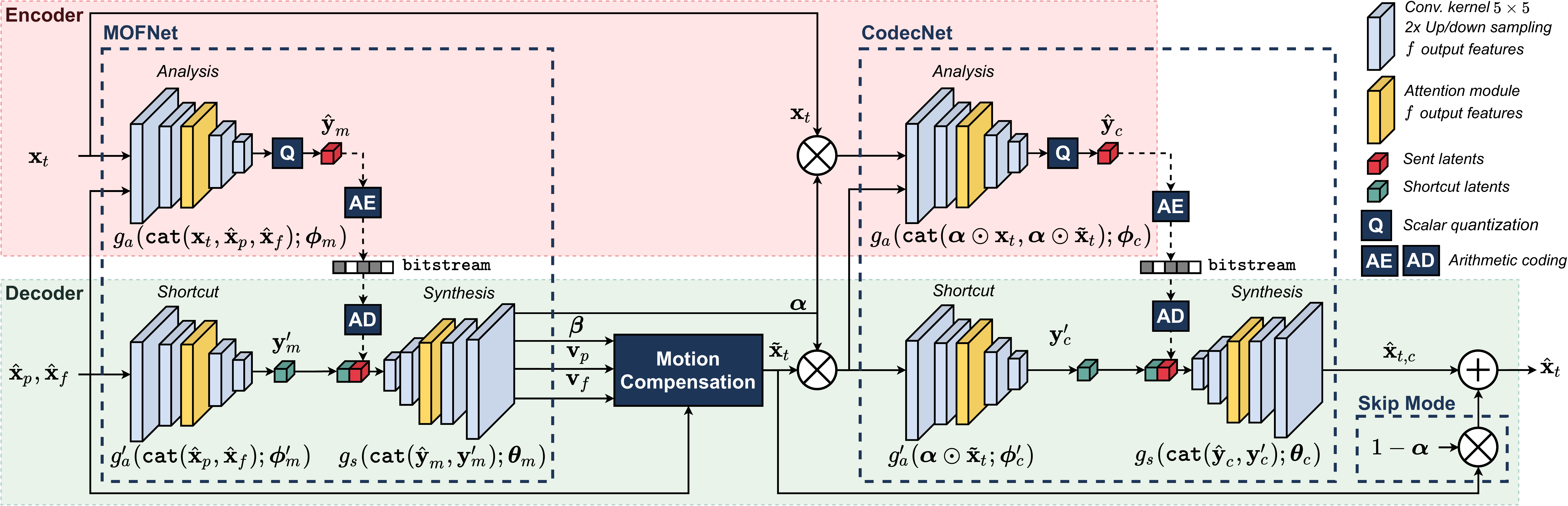}
    \caption{Diagram of the system. A detailed version can be found in appendix
    \ref{app:detailarchitecture}. Arithmetic coding uses hyperpriors
    \citep{DBLP:conf/iclr/BalleMSHJ18} omitted for clarity. Attention modules
    are implemented as proposed by \citet{cheng2020learned} and $f = 128$. There
    are 20 millions learnable parameters $\left\{\boldsymbol{\phi},\boldsymbol{\theta}\right\}$.}
    \label{fig:overalldiagram}
\end{figure}

Conditional coding \citep{LaduneMMSP20} allows to exploit decoder-side information
more efficiently than residual coding. Its architecture is similar to an
auto-encoder \citep{DBLP:conf/iclr/BalleMSHJ18}, with one additional
\textit{shortcut} transform (Fig. \ref{fig:overalldiagram}). It
can be understood through the description of its 3 transforms.\\
\textbf{Shortcut transform} $g^\prime_a$ (\textit{Decoder})---Its role is to extract information
from the reference frames available at the decoder (\textit{i.e.} at no rate). The
information is computed as latents $\mathbf{y}^\prime$.\\
\textbf{Analysis transform} $g_a$ (\textit{Encoder})---It estimates and conveys
the information not available at the decoder \textit{i.e.} the unpredictable
part. The information is computed as latents $\hat{\mathbf{y}}$.\\
\textbf{Synthesis transform} $g_s$ (\textit{Decoder})---Latents from the analysis and shortcut
transforms are concatenated and synthesized to obtain the desired output.

Unlike residual coding, conditional coding leverages decoder-side information in
the latent domain. As noted by \citet{DBLP:conf/iccv/DjelouahCSS19}, this makes
the system more resilient to the absence of information at the decoder
(\textit{i.e.} for I-frames). Thus, MOFNet and CodecNet implement
conditional coding to be able to process I, P and B-frames as well as lowering
their rate. I and P-frames are compressed using the B-frames coding scheme, with
the same parameters, and ignore the unavailable elements.\\
\textbf{I-frame}---Motion compensation is not available. As such,
MOFNet is ignored, $\balph$ is set to 1 and CodecNet conveys the whole frame, with its shortcut latents
$\mathbf{y}^\prime_c$ set to $0$.\\
\textbf{P-frame}---Bi-directional motion compensation is not available. $\bbeta$
is set to 1 to only rely on the prediction from $\pastref$. MOFNet shortcut
latents $\mathbf{y}^\prime_m$ are set to $0$.

\section{Training}

The training aims at learning to code I, P and B-frames. As such, it considers
the smallest coding configuration featuring all 3 types of frame: a GOP of size
2 plus the preceding I-frame. Each training iteration consists in the coding of
the 3 frames, followed by a single back-propagation to minimize the
rate-distortion cost of \eqref{eq:loss}. Unlike previous works, the entire
learning process is achieved through this rate-distortion loss. No element of
the system requires a pre-training or a dedicated loss term. Moreover, coding the
entire GOP in the forward pass enables the system to model the dependencies
between coded frames, leading to better coding performance.

The training set is made of 400~000 videos crops of size $256 \times 256$, with
various resolutions (from 540p to 4K) and framerates (from 24 to 120 fps). The
original videos are from several datasets: KonViD-1k \citep{hosu2017konstanz},
CLIC20 P-frame and Youtube-NT \citep{yang2020hierarchical}. The batch size is 4
and the learning rate is set to $10^{-4}$ and decreased to $10^{-5}$ during the
last epochs. Rate-distortion curves are obtained by training systems for
different $\lambda$.

\section{Visual Illustrations}
\label{sec:visualization}

\begin{figure}[htb]
    \centering 
\begin{subfigure}[t]{0.24\textwidth}
  \includegraphics[width=\linewidth]{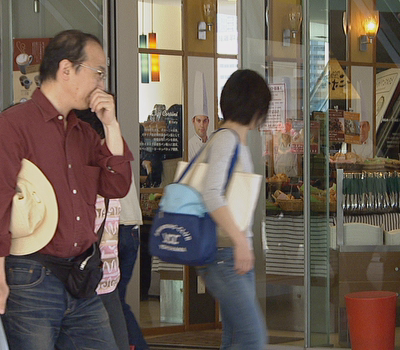}
  \caption{Frame to code $\mathbf{x}_t$}
  \label{subfig:code}
\end{subfigure}\hfil 
\begin{subfigure}[t]{0.265\textwidth}
    \includegraphics[width=\linewidth]{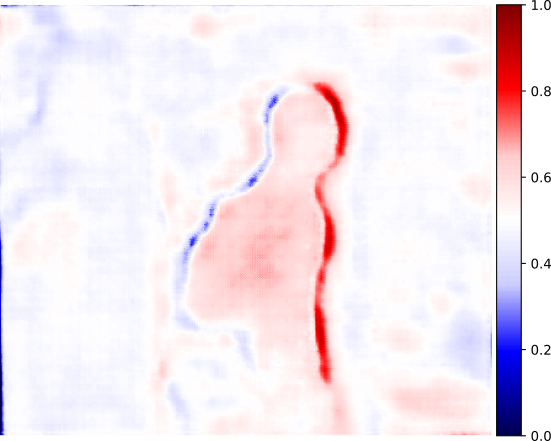}
    \caption{Prediction weighting $\bbeta$}
    \label{subfig:beta}
  \end{subfigure}\hfil 
  \begin{subfigure}[t]{0.24\textwidth}
    \includegraphics[width=\linewidth]{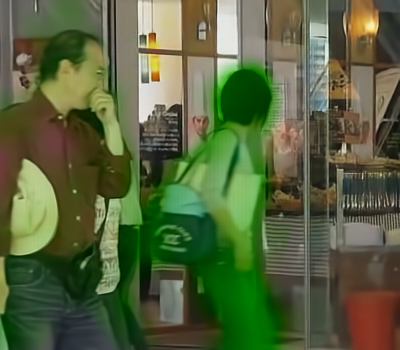}
    \caption{Skip mode $(1-\balph) \odot \pred$}
    \label{subfig:skipmode}
    \end{subfigure}

    \vspace{-0.1cm}
    \begin{subfigure}[t]{0.23\textwidth}
        \includegraphics[width=\linewidth]{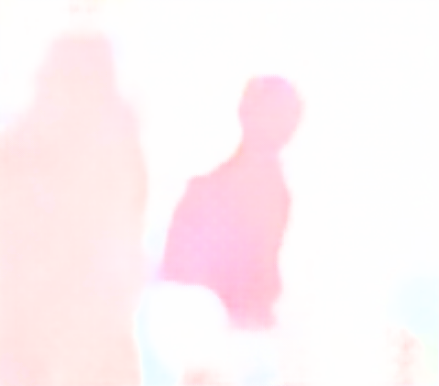}
        \caption{Optical flow $\futureflow$}
        \label{subfig:flow}
    \end{subfigure}\hfil
    \begin{subfigure}[t]{0.23\textwidth}
        \includegraphics[width=\linewidth]{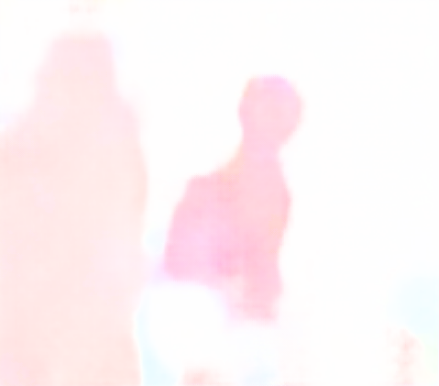}
        \caption{$\futureflow$ from $g_s(\mathbf{y}^\prime_m; \boldsymbol{\theta}_m)$}
        \label{subfig:flowshortcut}
      \end{subfigure}\hfil 
      \begin{subfigure}[t]{0.23\textwidth}
        \includegraphics[width=\linewidth]{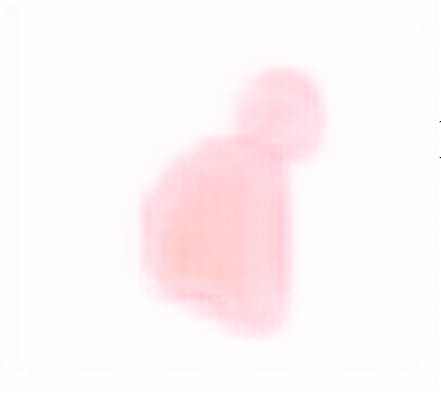}
        \caption{$\futureflow$ from $g_s(\hat{\mathbf{y}}_m; \boldsymbol{\theta}_m)$}
        \label{subfig:flowsent}
      \end{subfigure}
\caption{B-frame coding from the \textit{BQMall} sequence featuring
moving people on a static background. This crop PSNR is $31.57$ dB, MOFNet
rate is $322$ bits and CodecNet rate is $2~240$ bits. Second
row shows $\futureflow$ computed by MOFNet synthesis transform from both latents
$\texttt{cat}(\hat{\mathbf{y}}_m,\mathbf{y}^\prime_m)$, from shortcut latents
$\mathbf{y}^\prime_m$ and from the transmitted latent $\hat{\mathbf{y}}_m$.}
\end{figure}

This section shows the different quantities at stakes when coding a B-frame
$\mathbf{x}_t$ (Fig. \ref{subfig:code}). First, MOFNet outputs two optical
flows $(\pastflow,\futureflow)$ (Fig. \ref{subfig:flow}), the prediction
weighting $\bbeta$ (Fig. \ref{subfig:beta}) and the coding mode selection
$\balph$. The temporal prediction is then computed following \eqref{eq:pred}.
Most of the time, $\bbeta \simeq 0.5$, mitigating the noise from both bilinear
warpings. When the background is disoccluded by a moving object (\textit{e.g.}
the woman), $\bbeta$ equals $0$ on one side of the object and $1$ on the other
side. This allows to retrieve the background from where it is available. The
competition between Skip mode and CodecNet is weighted by $\balph$. Here, most
of $\hat{\mathbf{x}}_t$ comes from the Skip mode\footnote{Video frames are in
YUV format. Thus zeroed areas appear green.} (Fig. \ref{subfig:skipmode}).
However, the less predictable parts, \textit{e.g.} the woman, are sent by
CodecNet.

To illustrate the conditional coding, $\futureflow$ is computed by the MOFNet
synthesis transform using only the shortcut latents $\mathbf{y}^\prime_m$ (Fig.
\ref{subfig:flowshortcut}), the transmitted ones $\hat{\mathbf{y}}_m$ (Fig.
\ref{subfig:flowsent}) or both (Fig. \ref{subfig:flow}). The shortcut transform
captures the nature of the motion in $\mathbf{y}^\prime_m$, which allows to
synthesize most of $\futureflow$ without any transmission involved. In contrast,
$\hat{\mathbf{y}}_m$ consists in a refinement of the flow magnitude. The
rate of $\hat{\mathbf{y}}_m$ is reduced by using a low spatial resolution,
unlike $\mathbf{y}^\prime_m$ which keeps all the spatial accuracy.

\newcommand{\behaviorroot}{\imagepath/}

\section{Rate-Distortion Results}

The proposed system is assessed against \texttt{x265}\footnote{Preset medium,
the exact command line can be found in appendix \ref{subsec:seqbyseqbd}.}, an
implementation of HEVC. The quality is measured with the PSNR and the BD-rate
\citep{Bjontegaard} indicates the rate difference for the same distortion
between two coders. The test sequences are from the HEVC Common Test Conditions
\citep{HEVC_CTC}. The system flexibility is tested under three coding
configurations: All Intra (AI) \textit{i.e.} coding only the first I-frame,
Low-delay P (LDP) \textit{i.e.} coding one I-frame plus 8 P-frames and Random
Access (RA) \textit{i.e.} coding one I-frame plus a GOP of size 8. BD-rates of
the proposed coder against HEVC are presented in the Table \ref{table:bdrate}.

\begin{table}[h]
    \centering
    \caption{BD-rate of the proposed coder against HEVC.
    Negative results indicate that the proposed coder requires less rate than HEVC for equivalent quality.
    }
    \begin{tabular}{l||rrrrr|r}
        \multirow{2}{*}{Coding configuration}              & \multicolumn{5}{c|}{Class (Resolution)}                                & \multirow{2}{*}{Average}\\
                            &    A (1600p)       &    B (1080p)      &     C (480p)       &     D (240p)       &     E (720p)       &                         \\
        \hline 
        All Intra (AI)      & $\mathbf{-11.3}$\% & $\mathbf{-9.6}$\% & $\mathbf{-14.8}$\% & $\mathbf{-45.6}$\% & $\mathbf{-25.8}$\% & $\mathbf{-21.4}$\%      \\
        Low-delay P (LDP)   & $\mathbf{-4.7}$\%  & $29.1$\%          & $14.3$\%           & $\mathbf{-9.5}$\%  & $10.0$\%           & $7.8$\%                 \\
        Random Access (RA)  & $5.3$\%            & $29.9$\%          & $7.0$\%            & $\mathbf{-27.2}$\% & $\mathbf{-18.7}$\% & $\mathbf{-0.7}$\%       \\
    \end{tabular}
\label{table:bdrate}
\end{table}

The proposed system outperforms HEVC in AI configuration, proving that it
properly handles I-frames. It is on par with HEVC for RA coding and slightly
worse than HEVC for LDP coding. This shows that the same coder is also able to
efficiently code P and B-frames, without affecting the I-frames performance. To
the best of our knowledge, this is the first system to achieve compelling
performance under different coding configurations with a single end-to-end
learned coder for the three types of frame.

\section{Conclusion}

This paper proposes a new framework for end-to-end video coding. It is based on
MOFNet and CodecNet, which use conditional coding to leverage the information
present at the decoder. Thanks to conditional coding, all types of frame (I, P
\& B) are processed using the same coder with the same parameters, offering a great
flexibility in the coding configuration. The entire training process is
performed through the minimization of a unique rate-distortion cost. Its flexibility is
illustrated under three coding configurations: All Intra, Low-delay P and Random
Access, where the system achieves performance competitive with HEVC.\\
The main focus of this work is not in the internal design of the networks architecture
(MOFNet and CodecNet). Future work will investigate more advanced architectures,
from the optical flow estimation or the learned image coding literature, which
should bring performance gains.

\newpage
\bibliographystyle{iclr2021_conference}
\bibliography{refs_wo_url}

\appendix
\section{Supplementary Rate-distortion Results}

\subsection{Sequence-by-sequence BD-rates}
\label{subsec:seqbyseqbd}
Table \ref{table:detailedbdrate} details the sequence-by-sequence BD-rates that
gives the averaged results presented in Table \ref{table:bdrate}. HEVC
compression is achieved using \texttt{ffmpeg} with the following command:

\begin{flushleft}
    \small
    \texttt{ffmpeg -video\_size WxH -i in.yuv -c:v libx265 -pix\_fmt yuv420p -x265-params
    "keyint=9:min-keyint=9" -crf QP -preset medium -tune psnr out.mp4}
\end{flushleft}

The BD-rate is computed using four quality factors \texttt{QP} $= \left\{27, 32,
37, 42\right\}$. The Low-delay P configuration is obtained by changing the
\texttt{tune} option to \texttt{zerolatency}. \texttt{WxH} denotes the video
width and height.

\newcommand{\no}[1]{\textcolor{nored}{#1}}
\newcommand{\yes}[1]{\textcolor{yesgreen}{#1}}

\begin{table}[h]
    \centering
    \caption{BD-rate of the proposed coder against HEVC.
    Negative results indicate that the proposed coder requires less rate than HEVC for equivalent quality.
    }
    \begin{tabular}{cl||rrr}
        Class                       & \multirow{2}{*}{Sequence name} & \multicolumn{3}{c}{Coding configuration}            \\
        (Resolution)                &                                &  All Intra (AI)          & Low-delay P (LDP)         & Random Access (RA) \\
        \hline
        \multirow{5}{*}{A (1600p)}  & Traffic                        & \yes{$-13.1$\%}          & \no{$12.8$\%}             & \no{$9.2$\%} \\
                                    & PeopleOnStreet                 & \yes{$-18.6$\%}          & \yes{$-20.4$\%}           & \yes{$-11.4$\%} \\ 
                                    & Nebuta                         & \yes{$-2.6$\% }          & \yes{$-18.7$\%}           & \no{$16.8$\%} \\
                                    & SteamLocomotive                & \yes{$-10.8$\%}          & \no{$7.7$\%}              & \no{$6.7$\% }\\
        \multicolumn{2}{c||}{\textbf{Average}}                       & \yes{$\mathbf{-11.3}$\%} & \yes{$\mathbf{-4.7}$\%}   & \no{$\mathbf{5.3}$\%} \\ 
        \hline
        \multirow{6}{*}{B (1080p)}  & Kimono                         & \yes{$-28.7$\%}          & \yes{$-1.8$\%}            & \no{$18.5$\%}\\  
                                    & ParkScene                      & \yes{$-17.0$\%}          & \no{$3.4$\%}              & \no{$4.5$\%}\\  
                                    & Cactus                         & \yes{$-4.5$\% }          & \no{$6.3$\%}              & \no{$12.7$\%}\\  
                                    & BQTerrace                      & \yes{$-6.4$\% }          & \no{$86.9$\%}             & \no{$30.6$\%}\\  
                                    & BasketballDrive                & \yes{$-4.0$\% }          & \no{$50.6$\%}             & \no{$83.0$\%}\\    
        \multicolumn{2}{c||}{\textbf{Average}}                       & \yes{$\mathbf{-9.6}$\% } & \no{$\mathbf{29.1}$\%}    & \no{$\mathbf{29.9}$\%}\\    
        \hline
        \multirow{5}{*}{C (480p)}   & RaceHorses                     & \yes{$-22.7$\%}          & \yes{$-12.0$\%}           & \no{$16.8$\%}\\
                                    & BQMall                         & \yes{$-15.7$\%}          & \no{$20.3$\%}             & \yes{$-2.6$\%}\\
                                    & PartyScene                     & \yes{$-4.8$\% }          & \no{$38.6$\%}             & \no{$20.0$\%}\\
                                    & BasketballDrill                & \yes{$-25.6$\%}          & \no{$10.4$\%}             & \no{$6.1$\%}\\
        \multicolumn{2}{c||}{\textbf{Average}}                       & \yes{$\mathbf{-14.8}$\%} & \no{$\mathbf{14.3}$\%}    & \no{$\mathbf{7.0}$\%}\\
        \hline
        \multirow{5}{*}{D (240p)}   & RaceHorses                     & \yes{$-50.1$\%}          & \yes{$-26.0$\%}           & \yes{$-12.1$\%}\\
                                    & BQSquare                       & \yes{$-25.2$\%}          & \no{$29.8$\%}             & \yes{$-22.4$\%}\\
                                    & BlowingBubbles                 & \yes{$-49.4$\%}          & \yes{$-22.3$\%}           & \yes{$-33.4$\%}\\
                                    & BasketballPass                 & \yes{$-57.5$\%}          & \yes{$-19.5$\%}           & \yes{$-41.1$\%}\\
        \multicolumn{2}{c||}{\textbf{Average}}                       & \yes{$\mathbf{-45.6}$\%} & \yes{$\mathbf{-9.5}$\% }  & \yes{$\mathbf{-27.2}$\%}\\
        \hline
        \multirow{4}{*}{E (720p)}   & FourPeople                     & \yes{$-25.5$\%}          & \no{$3.8$\%}              & \yes{$-20.3$\%}\\
                                    & Johnny                         & \yes{$-25.2$\%}          & \no{$15.8$\%}             & \yes{$-18.5$\%}\\  
                                    & KristenAndSara                 & \yes{$-26.6$\%}          & \no{$10.5$\%}             & \yes{$-17.3$\%}\\
        \multicolumn{2}{c||}{\textbf{Average}}                       & \yes{$\mathbf{-25.8}$\%} & \no{$\mathbf{10.0}$\%}    & \yes{$\mathbf{-18.7}$\%}\\
        \hline
        \hline
    \multicolumn{2}{c||}{\textbf{All classes average}}               & \yes{$\mathbf{-21.4}$\%} & \no{$\mathbf{7.8}$\%}     & \yes{$\mathbf{-0.7}$\%}\\
    \end{tabular}
\label{table:detailedbdrate}
\end{table}

\newpage
\subsection{Supplementary Anchors}

Previous work \citep{DBLP:conf/cvpr/LuO0ZCG19,DBLP:conf/iccv/DjelouahCSS19} uses
AVC as an anchor. Table \ref{table:detailedbdrateavc} displays the BD-rate of the proposed system
against \texttt{x264} through the following command:

\begin{flushleft}
    \small
    \texttt{ffmpeg -video\_size WxH -i in.yuv -c:v libx264 -pix\_fmt yuv420p -g 9 -crf QP -preset medium -tune psnr out.mp4}
\end{flushleft}

The proposed system consistently outperforms AVC in all classes under the three coding configurations.


\begin{table}[h]
    \centering
    \caption{BD-rate of the proposed coder against AVC.
    Negative results indicate that the proposed coder requires less rate than AVC for equivalent quality.
    }
    \begin{tabular}{cl||rrr}
        Class                       & \multirow{2}{*}{Sequence name} & \multicolumn{3}{c}{Coding configuration}            \\
        (Resolution)                &                                &  All Intra (AI)          & Low-delay P (LDP)         & Random Access (RA) \\
        \hline
        \multirow{5}{*}{A (1600p)}  & Traffic                        & \yes{$-29.8$\%}          & \yes{$-13.5$\%}           & \yes{$-16.6$\%} \\
                                    & PeopleOnStreet                 & \yes{$-34.4$\%}          & \yes{$-32.6$\%}           & \yes{$-22.4$\%} \\ 
                                    & Nebuta                         & \yes{$-25.3$\%}          & \yes{$-53.2$\%}           & \yes{$-17.3$\%} \\
                                    & SteamLocomotive                & \yes{$-27.4$\%}          & \yes{$-12.5$\%}           & \yes{$-16.4$\% }\\
        \multicolumn{2}{c||}{\textbf{Average}}                       & \yes{$\mathbf{-29.2}$\%} & \yes{$\mathbf{-27.9}$\%} & \yes{$\mathbf{-18.1}$\%} \\ 
        \hline
        \multirow{6}{*}{B (1080p)}  & Kimono                         & \yes{$-40.8$\%}          & \yes{$-31.3$\%}           & \yes{$-21.0$\%}\\  
                                    & ParkScene                      & \yes{$-26.3$\%}          & \yes{$-10.2$\%}           & \yes{$-15.9$\%}\\  
                                    & Cactus                         & \yes{$-22.3$\%}          & \yes{$-15.0$\%}           & \yes{$-13.5$\%}\\  
                                    & BQTerrace                      & \yes{$-16.7$\%}          & \no{$32.2$\%}             & \yes{$-4.8$\%}\\  
                                    & BasketballDrive                & \yes{$-18.7$\%}          & \yes{$-0.8$\%}            & \no{$13.2$\%}\\    
        \multicolumn{2}{c||}{\textbf{Average}}                       & \yes{$\mathbf{-25.0}$\%} & \yes{$\mathbf{-5.0}$\%}   & \yes{$\mathbf{-8.4}$\%}\\    
        \hline
        \multirow{5}{*}{C (480p)}   & RaceHorses                     & \yes{$-22.4$\%}          & \yes{$-13.2$\%}           & \no{$9.5$\%}\\
                                    & BQMall                         & \yes{$-11.2$\%}          & \no{$8.2$\%}              & \yes{$-7.0$\%}\\
                                    & PartyScene                     & \no{$3.2$\%}             & \no{$28.7$\%}             & \no{$14.3$\%}\\
                                    & BasketballDrill                & \yes{$-28.9$\%}          & \yes{$-19.1$\%}           & \yes{$-21.7$\%}\\
        \multicolumn{2}{c||}{\textbf{Average}}                       & \yes{$\mathbf{-14.8}$\%} & \no{$\mathbf{1.1}$\%}     & \yes{$\mathbf{-1.2}$\%}\\
        \hline
        \multirow{5}{*}{D (240p)}   & RaceHorses                     & \yes{$-32.9$\%}          & \yes{$-18.2$\%}           & \yes{$-5.6$\%}\\
                                    & BQSquare                       & \yes{$-2.2$\%}           & \no{$42.1$\%}             & \yes{$-8.8$\%}\\
                                    & BlowingBubbles                 & \yes{$-26.9$\%}          & \yes{$-6.1$\%}            & \yes{$-21.5$\%}\\
                                    & BasketballPass                 & \yes{$-32.3$\%}          & \yes{$-8.3$\%}            & \yes{$-29.1$\%}\\
        \multicolumn{2}{c||}{\textbf{Average}}                       & \yes{$\mathbf{-23.6}$\%} & \no{$\mathbf{2.4}$\% }    & \yes{$\mathbf{-15.8}$\%}\\
        \hline
        \multirow{4}{*}{E (720p)}   & FourPeople                     & \yes{$-28.5$\%}          & \yes{$-14.7$\%}           & \yes{$-31.5$\%}\\
                                    & Johnny                         & \yes{$-27.5$\%}          & \yes{$-7.5$\%}            & \yes{$-32.8$\%}\\  
                                    & KristenAndSara                 & \yes{$-29.5$\%}          & \yes{$-15.5$\%}           & \yes{$-32.5$\%}\\
        \multicolumn{2}{c||}{\textbf{Average}}                       & \yes{$\mathbf{-28.5}$\%} & \yes{$\mathbf{-12.5}$\%}  & \yes{$\mathbf{-32.2}$\%}\\
        \hline
        \hline
    \multicolumn{2}{c||}{\textbf{All classes average}}               & \yes{$\mathbf{-24.2}$\%} & \yes{$\mathbf{-8.4}$\%}   & \yes{$\mathbf{-15.1}$\%}\\
    \end{tabular}
\label{table:detailedbdrateavc}
\end{table}

\newpage
\section{System Behavior on the \textit{FourPeople} Sequence}

Additional illustrations of the system behavior are given here for the
\textit{FourPeople} sequence, extracted from the HEVC Common Test Conditions. This
sequence shows four people slightly moving in front of a still background. The
first frame is coded as an I-frame and the next 8 frames are compressed using a
(random access) GOP of size 8.

\subsection{Rate-distortion curves}

\begin{figure}[H]
    \centering
    \begin{tikzpicture}
        \begin{axis}[
            grid= both ,
            xlabel = {Rate (Mbit/s)} ,
            ylabel = {PSNR (dB)} ,
            ylabel near ticks,
            xlabel near ticks,
            x tick label style={
                /pgf/number format/.cd,
                fixed,
                precision=2
                },
            xtick distance={0.5},
            ytick distance={2},
            minor y tick num=1,
            minor x tick num=1,
            xmin=0.25,xmax=2.25,ymin=32,ymax=40,
            title={Rate-distortion curves on \textit{FourPeople} --- BD-rate is -20.3\% w.r.t. HEVC},
            legend style={at={(0.0,1.0)},anchor=north west}
        ]

        \addplot[solid, thick, mark=triangle*, palet_dark_blue, only marks, forget plot] coordinates{
            (0.76186667,32.66859018)
            (1.21962667,35.54858311)
            (1.95802667,38.3934803)
            (3.13626667,40.88808664)
            };
        \addplot[thick, domain=0.76186667:3.13626667,palet_dark_blue] {5.8289*ln(x) + 34.3369};
        \addlegendentry{AVC};

        \addplot[thick, solid, mark=square*, emerald,only marks, forget plot] coordinates {
            (0.7,32.73862316)
            (1.07573333,35.64802191)
            (1.2735,36.939)
            (1.65813333,38.46134672)
            (2.59253333,40.99592513)
        };
        \addplot[thick, domain=0.7:2.59253333,emerald] {6.7031*ln(x) + 35.16955};
        \addlegendentry{HEVC};

        \addplot[thick, solid, palet_red, mark=*,only marks, forget plot] coordinates {
            (0.6027264,33.879)
            (0.8515584,35.583)
            (1.2109823999999998,37.29)
            (1.6367616,38.887)
        };
        \addplot[thick, domain=0.6027264:1.6367616,palet_red] {4.9924*ln(x) + 36.3883};
        \addlegendentry{Ours};

        \node (mark) [thick, draw, black, circle, minimum size = 10pt, inner sep=4pt, ultra thick]
            at (axis cs: 1.2109823999999998, 37.29) [] {};
            \node (mark) [thick, draw, black, circle, minimum size = 10pt, inner sep=4pt, ultra thick]
            at (axis cs: 1.2735,36.939) [] {};
        
        \node at (axis cs: 1.2109823999999998, 37.29) [left=0.25cm] {\textit{Visual examples}};

        \end{axis}
    \end{tikzpicture}
    \caption{Rate-distortion curve of the proposed system against AVC and HEVC
    for the \textit{FourPeople} sequence. The circled points are
    used to generate the visual examples.}
    \label{fig:fourpeoplerd}
\end{figure}
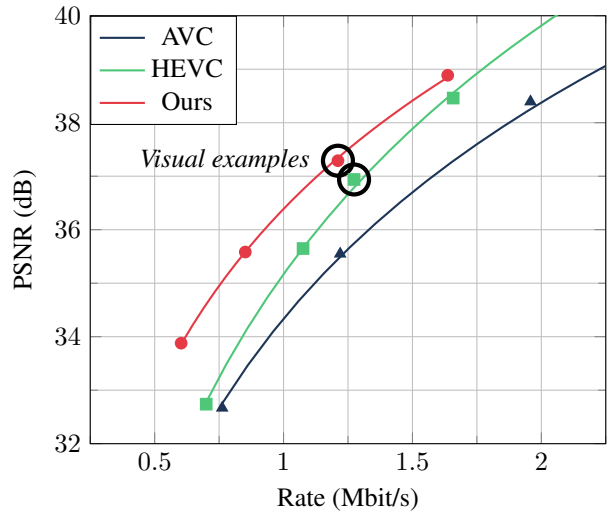

Rate-distortion results of the proposed coder against HEVC and AVC are presented
in Figure \ref{fig:fourpeoplerd}. On this sequence, the system significantly
outperforms HEVC across the entire rate range.

\newpage
\subsection{Rate Distribution inside a GOP}

\begin{figure}[H]
    \centering
    \begin{subfigure}{\textwidth}
        \pgfplotsset{compat=1.6}
        \begin{tikzpicture}
            \begin{axis}[
                height=4.5cm,
                width=\linewidth,
                ylabel=Rate (Mbit/s),
                xlabel=Frame index and type,
                xmin=0, xmax=9.00,
                ymin=0, ymax=12.00,
                ytick distance=3,
                yminorgrids=true,
                ymajorgrids=true,
                minor y tick num=1,
                ybar, bar width=15,
                enlarge x limits=0.1,
                xticklabels={I0,B1,B2,B3,B4,B5,B6,B7,P8, Avg},
                xtick={0,...,9},
                xtick pos=left,
                ytick pos=left,
                nodes near coords,
                nodes near coords align={vertical},
            ]
            \addplot[ybar, fill=palet_medium_blue] coordinates {
                    (0,8.9)
                    (1,0.1)
                    (2,0.1)
                    (3,0.1)
                    (4,0.3)
                    (5,0.1)
                    (6,0.2)
                    (7,0.1)
                    (8,1.9)
                    (9,1.2)
                };
            \end{axis}
        \end{tikzpicture}
    \caption{Rate per frame in a GOP.}
    \end{subfigure}\hfil 

    \medskip
    \begin{subfigure}{\textwidth}
        \pgfplotsset{compat=1.6}
        \begin{tikzpicture}
            \begin{axis}[
                height=4.5cm,
                width=\linewidth,
                ylabel=PSNR (dB),
                xlabel=Frame index and type,
                xmin=0, xmax=9.00,
                ymin=35, ymax=40,
                ytick distance=1,
                yminorgrids=true,
                ymajorgrids=true,
                minor y tick num=0,
                ybar, bar width=15,
                enlarge x limits=0.1,
                xticklabels={I0,B1,B2,B3,B4,B5,B6,B7,P8, Avg},
                xtick={0,...,9},
                xtick pos=left,
                ytick pos=left,
                nodes near coords,
                nodes near coords align={vertical},
            ]
            \addplot[ybar, fill=palet_medium_blue] coordinates {
                    (0,37.9)
                    (1,37.5)
                    (2,37.4)
                    (3,37.2)
                    (4,37.3)
                    (5,37.1)
                    (6,37.1)
                    (7,37.1)
                    (8,37.1)
                    (9,37.3)
                };
            \end{axis}
        \end{tikzpicture}
    \caption{PSNR per frame in a GOP.}
    \end{subfigure}
    \caption{Distribution of the rate and the PSNR across all frames of a GOP.
    \textit{Avg} denotes the mean value computed on I-frame and the GOP of size 8.}
    \label{fig:gopdetail4p}
\end{figure}
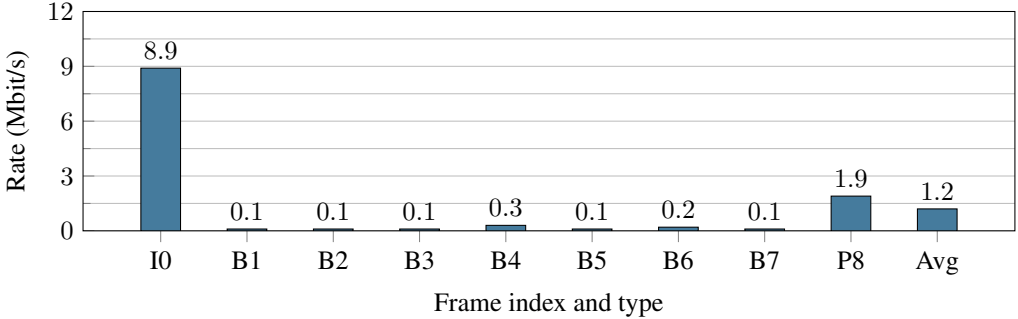
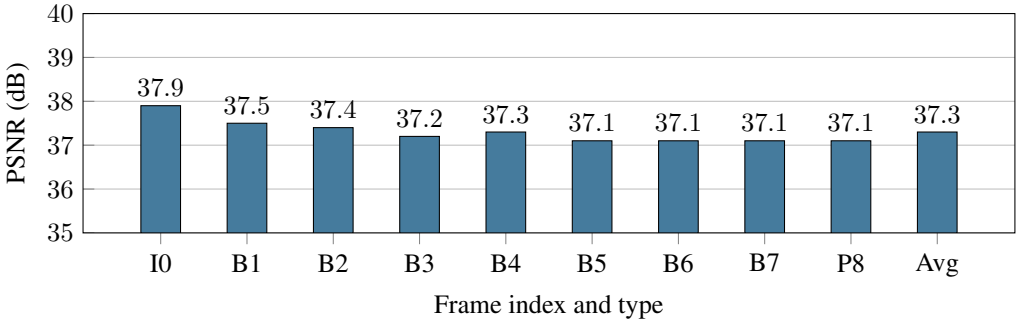

The Figure \ref{fig:gopdetail4p} exhibits the distribution of the rate and of
the PSNR across the frames of a GOP. The PSNR is stable for all the coded
frames, ensuring temporal consistency. As their distortion is roughly the same,
their rate is function of their \textit{predictability}. Indeed, the less
predictable is a frame, the more information are transmitted. As a result,
frames B4 or P8 require more bits to be sent than other inter-frames because
their references are temporally further, making them less predictable.

\subsection{Detailed B-frame Coding}

This section displays the quantities involved when coding a B-frame
$\mathbf{x}_t$ (Fig. \ref{fig:4poriginalframe}). The two optical flows
$\pastflow$, $\futureflow$ (Fig. \ref{fig:4ppastflow} and
\ref{fig:4pfutureflow}) and the pixel-wise bi-directional weighting $\bbeta$
(Fig. \ref{fig:4pbeta}) are used to compute the temporal prediction $\pred$ as
specified in \eqref{eq:pred}. Because both flows represent the motion
\textit{from} $\mathbf{x}_t$ \textit{to} a reference, they are activated at the
same spatial locations but with different directions, resulting in different
visualization colors. Some areas (\textit{e.g.} the left man's hand) exhibits a
motion which is not well captured by the system, causing checkerboard artifacts
in the visualization. Disocclusions occurring due to moving objects are handled using $\bbeta
= 0$ on one side of the objects and $\bbeta=1$ on the other side. This behavior
can be seen around the left man's arm. \\

MOFNet also computes and transmits the coding mode selection $\balph$ (Fig.
\ref{fig:4palpha}). Areas in blue ($\balph = 0$) rely on Skip mode to be
reconstructed (Fig. \ref{fig:4pskip}) while areas in red ($\balph = 1$) are
transmitted with CodecNet (Fig. \ref{fig:4pcodec}). Most of the decoded frame
(Fig. \ref{fig:4pcodedframe}) comes from Skip mode whereas areas transmitted
with CodecNet are only those not well predicted enough \textit{e.g.} the left
man's hand. Skip mode relevance is illustrated through the spatial distribution
of CodecNet rate (Fig. \ref{fig:4pcodecrate}). Thanks to Skip mode, only few
areas of the frame use CodecNet, resulting in few areas for which bits are
spent. Lastly, the spatial distribution of MOFNet rate (Fig.
\ref{fig:4pmofnetrate}) shows that all of the coding scheme side-information
$(\balph, \bbeta, \pastflow, \futureflow)$ are transmitted for a low rate.

\newpage
\newcommand{\detailBfourp}{\imagepath/}
\newcommand{\sizesubfig}{0.475\textwidth}
\begin{figure}[H]
    \centering
    \begin{subfigure}{\sizesubfig}
        \includegraphics[width=\linewidth]{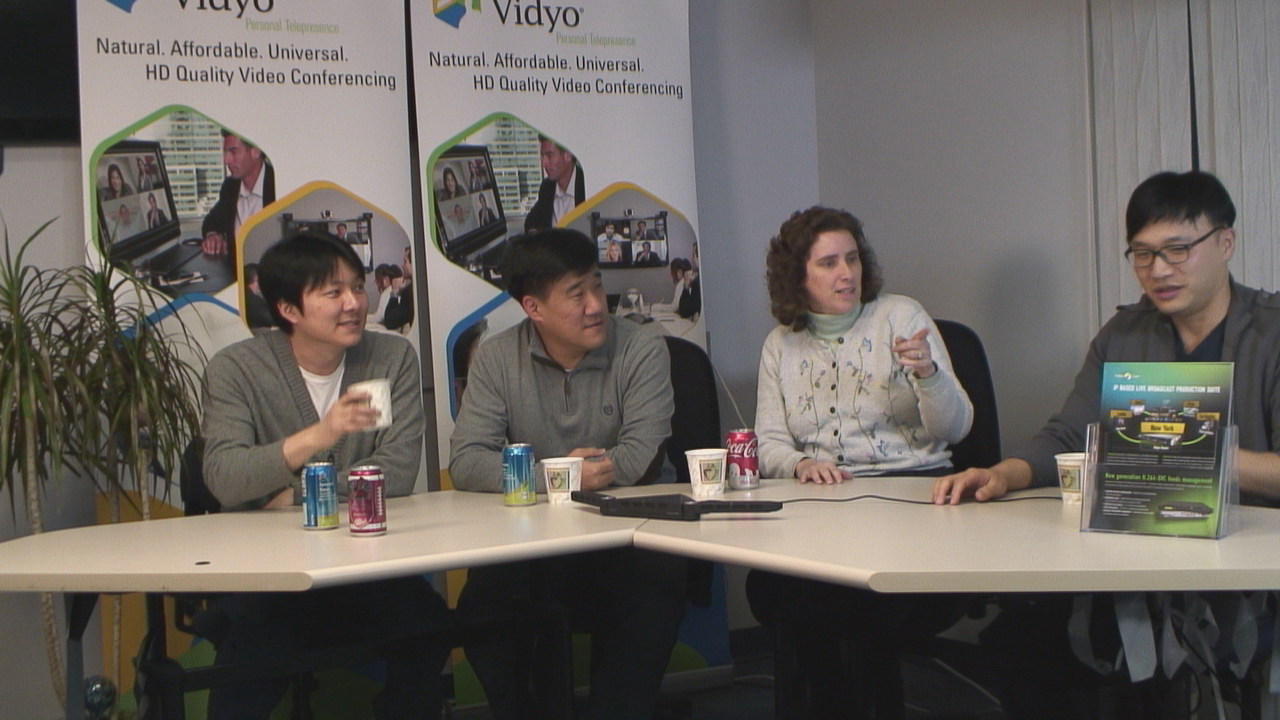}
        \caption{Frame to code $\mathbf{x}_t$}
        \label{fig:4poriginalframe}
    \end{subfigure}
    \begin{subfigure}{\sizesubfig}
        \includegraphics[width=\linewidth]{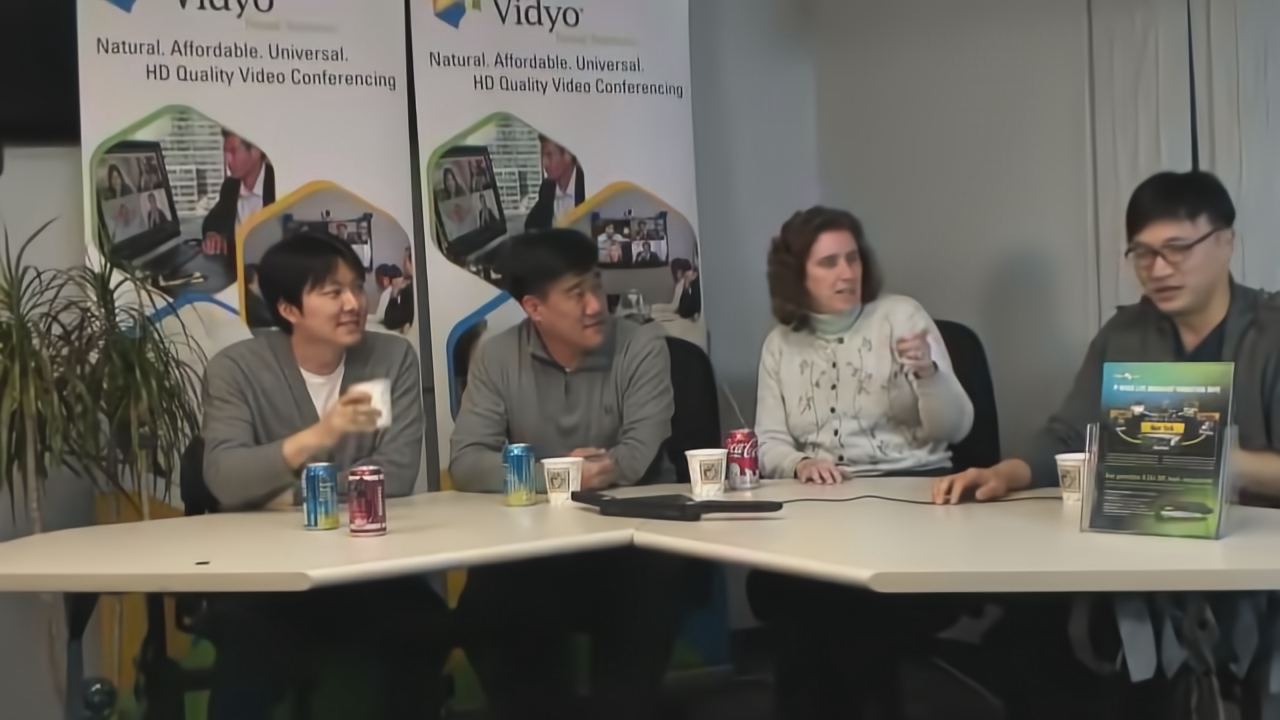}
        \caption{Compressed frame $\hat{\mathbf{x}}_t$, PSNR is $37.31$ dB}
        \label{fig:4pcodedframe}
    \end{subfigure}

    \medskip
    \begin{subfigure}{\sizesubfig}
        \includegraphics[width=\linewidth]{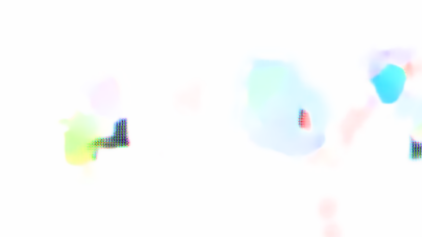}
        \caption{Optical flow $\pastflow$}
        \label{fig:4ppastflow}
    \end{subfigure}
    \begin{subfigure}{\sizesubfig}
        \includegraphics[width=\linewidth]{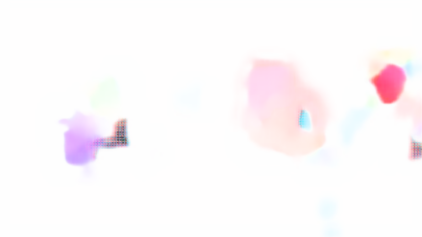}
        \caption{Optical flow $\futureflow$}
        \label{fig:4pfutureflow}
    \end{subfigure}

    \medskip
    \begin{subfigure}{\sizesubfig}
        \includegraphics[width=\linewidth]{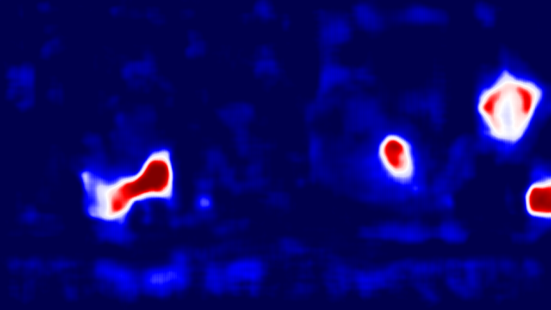}
        \caption{Coding mode $\balph$. Red: CodecNet, blue: Skip}
        \label{fig:4palpha}
    \end{subfigure}
    \begin{subfigure}{\sizesubfig}
        \includegraphics[width=\linewidth]{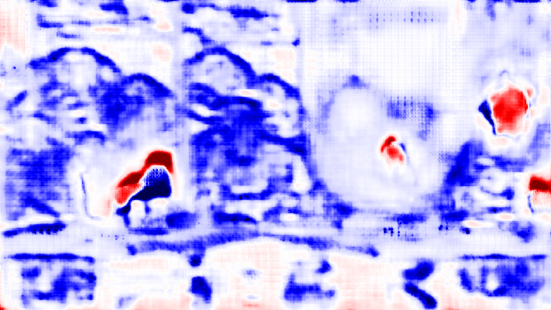}
        \caption{Bi-directional weighting $\bbeta$. Red $=1$, blue $=0$}
        \label{fig:4pbeta}
    \end{subfigure}

    \medskip
    \begin{subfigure}{\sizesubfig}
        \includegraphics[width=\linewidth]{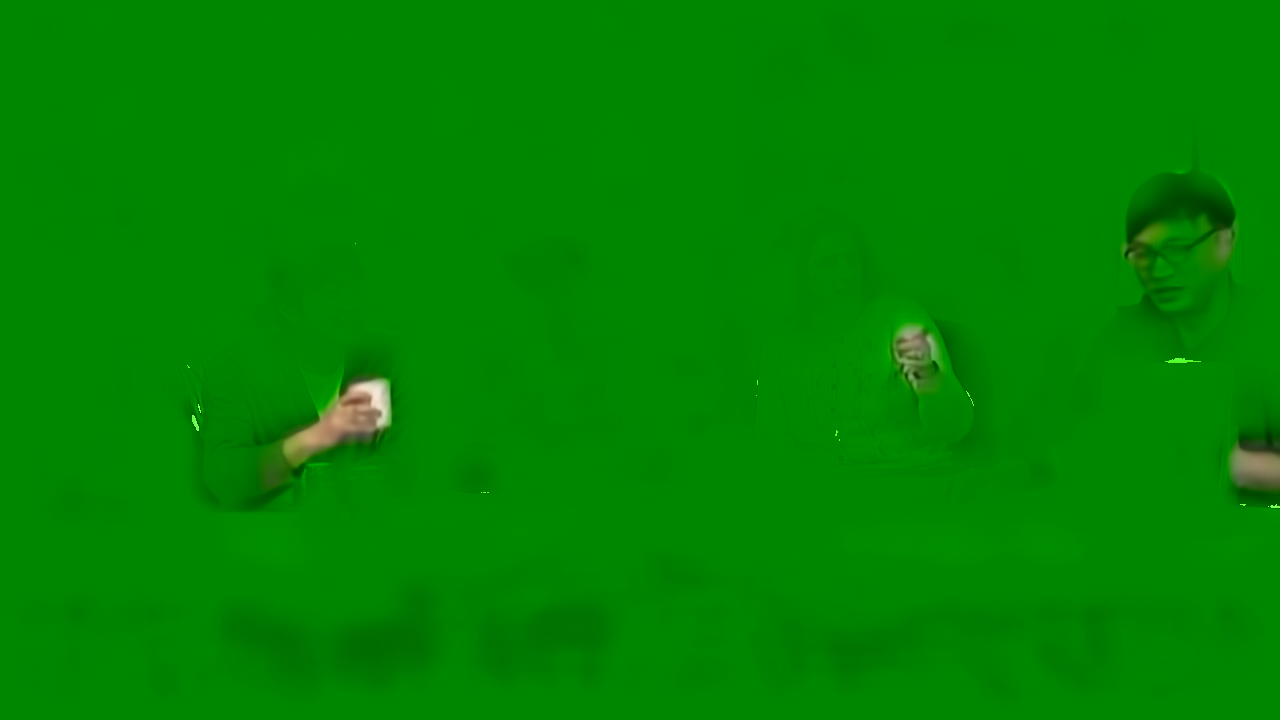}
        \caption{CodecNet part $c(\balph \odot \mathbf{x}_t, \balph \odot \pred)$}
        \label{fig:4pcodec}
    \end{subfigure}    
    \begin{subfigure}{\sizesubfig}
        \includegraphics[width=\linewidth]{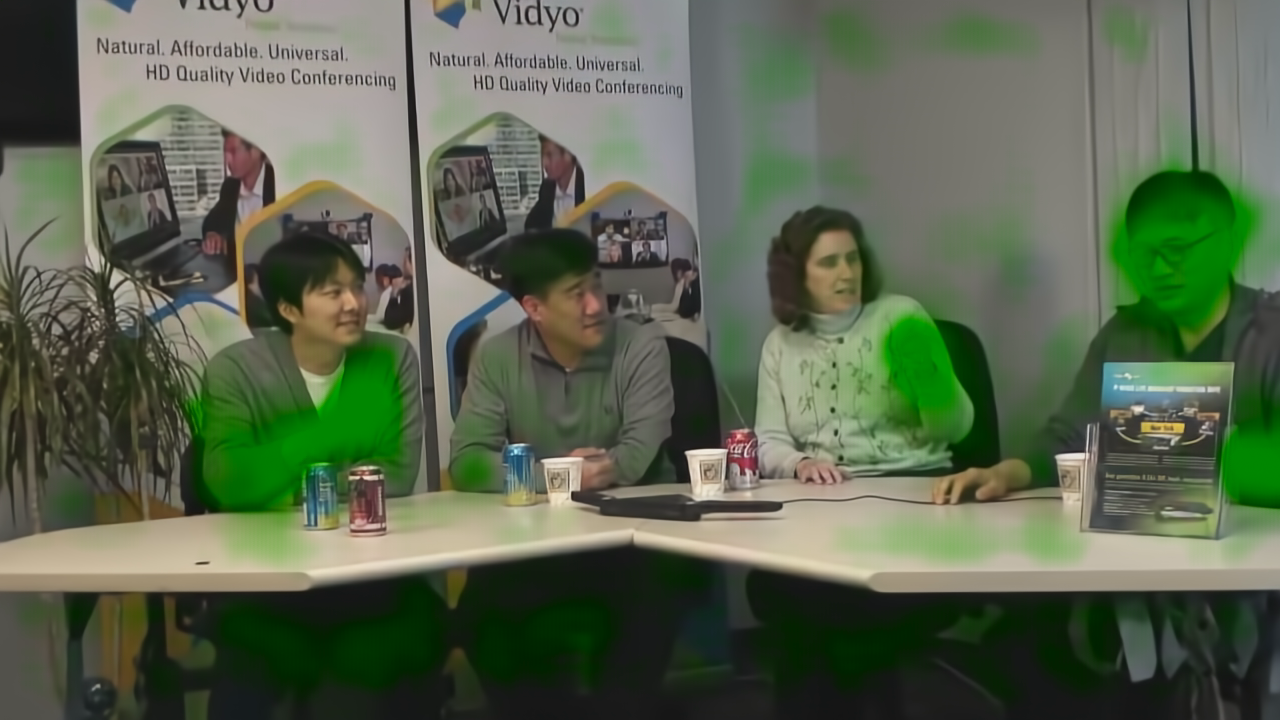}
        \caption{Skip part $(1-\balph ) \odot \pred$}
        \label{fig:4pskip}
    \end{subfigure}
    
    \medskip
    \begin{subfigure}{\sizesubfig}
        \includegraphics[width=\linewidth]{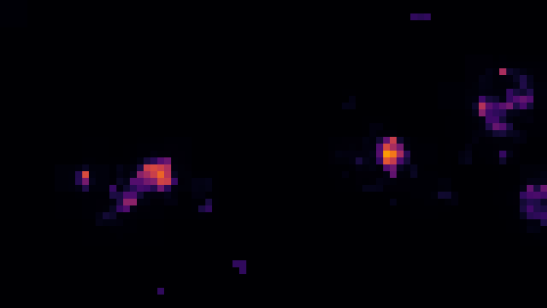}
        \caption{Distribution of CodecNet rate $R_c = 4~332$ bits}
        \label{fig:4pcodecrate}
    \end{subfigure}    
    \begin{subfigure}{\sizesubfig}
        \includegraphics[width=\linewidth]{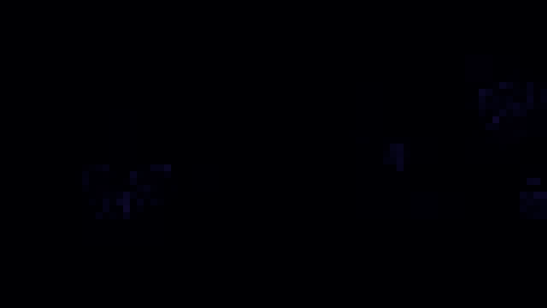}
        \caption{Distribution of MOFNet rate $R_m = 922$ bits}
        \label{fig:4pmofnetrate}
    \end{subfigure}
    \caption{Detailed visualizations for B-frame coding.}
    \label{fig:4pdetails}
\end{figure}

\subsection{Conditional Coding}

\begin{figure}[H]
    \centering
    \begin{subfigure}{\textwidth}
        \centering
        \begin{tikzpicture}[spy using outlines={circle,palet_red,magnification=5,size=3cm, ultra thick}]
            \node {\includegraphics[width=0.6\linewidth]{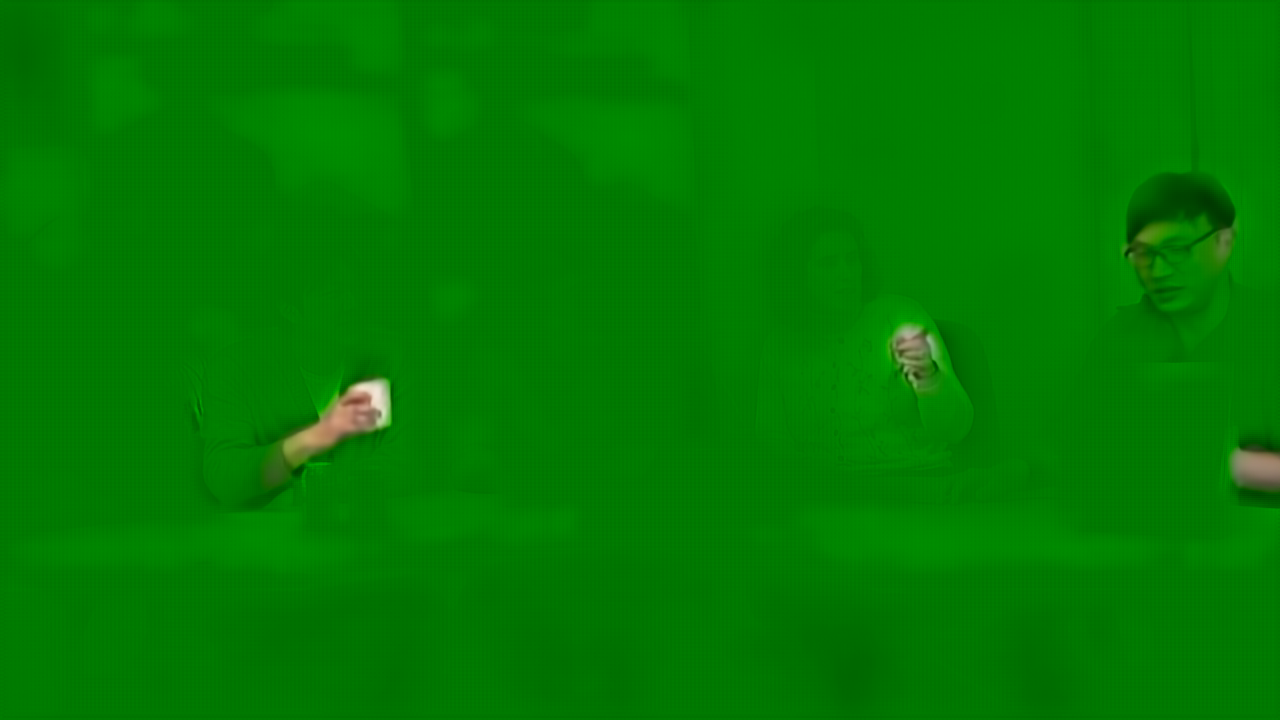}};
            \spy[size=3.5cm, connect spies] on (-1.8,-0.35) in node [left] at (-3.1,0);
            \spy[size=3.5cm, connect spies] on (1.8,0) in node [left] at (6.5,0);
        \end{tikzpicture}
        \caption{CodecNet output from both latents: $g_s(\texttt{cat}(\hat{\mathbf{y}}_c, \mathbf{y}_c^\prime); \boldsymbol{\theta}_c)$}
        \label{fig:4pcodecall}  
    \end{subfigure}

      \medskip
      \begin{subfigure}{\textwidth}
        \centering
        \begin{tikzpicture}[spy using outlines={circle,palet_red,magnification=5,size=3cm, connect spies, ultra thick}]
            \node {\includegraphics[width=0.6\linewidth]{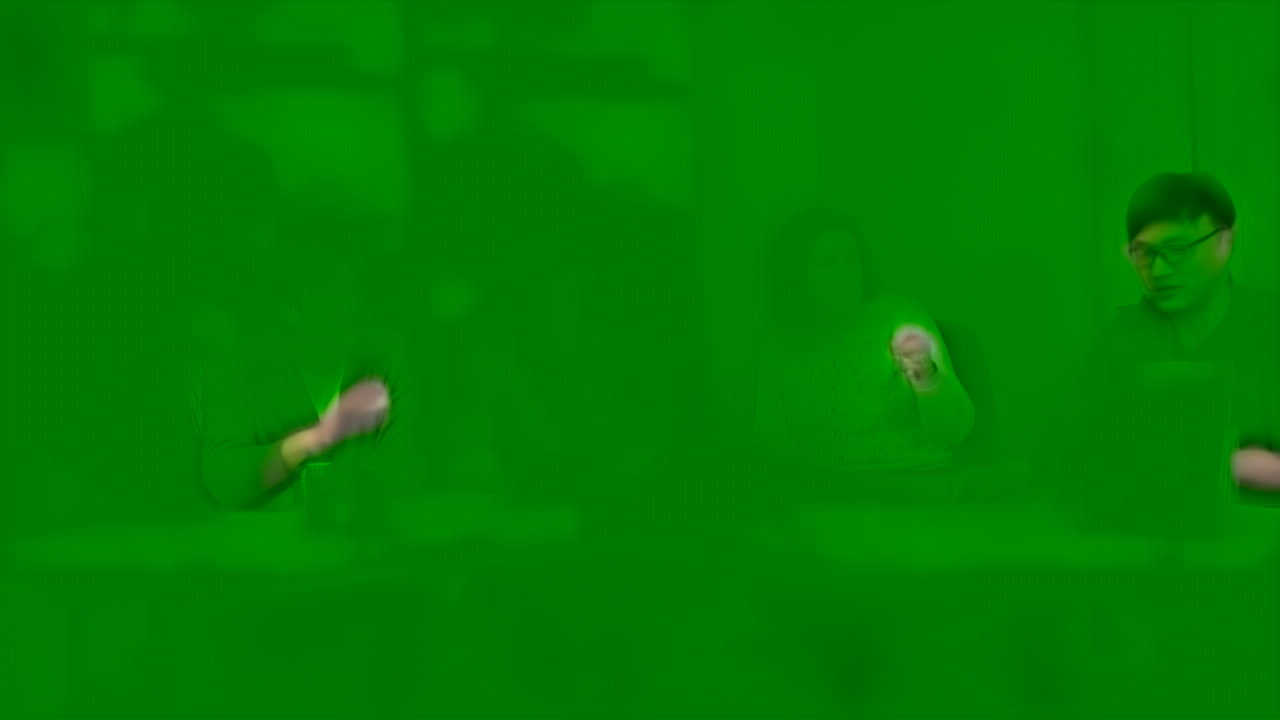}};
            \spy[size=3.5cm, connect spies] on (-1.8,-0.35) in node [left] at (-3.1,0);
            \spy[size=3.5cm, connect spies] on (1.8,0) in node [left] at (6.5,0);
        \end{tikzpicture}
        \caption{CodecNet output from shortcut latents only: $g_s(\mathbf{y}_c^\prime; \boldsymbol{\theta}_c)$}
        \label{fig:4pcodecshortcut}  
    \end{subfigure}

      \medskip
      \begin{subfigure}{\textwidth}
        \centering
        \begin{tikzpicture}[spy using outlines={circle,palet_red,magnification=5,size=3cm, connect spies, ultra thick}]
            \node {\includegraphics[width=0.6\linewidth]{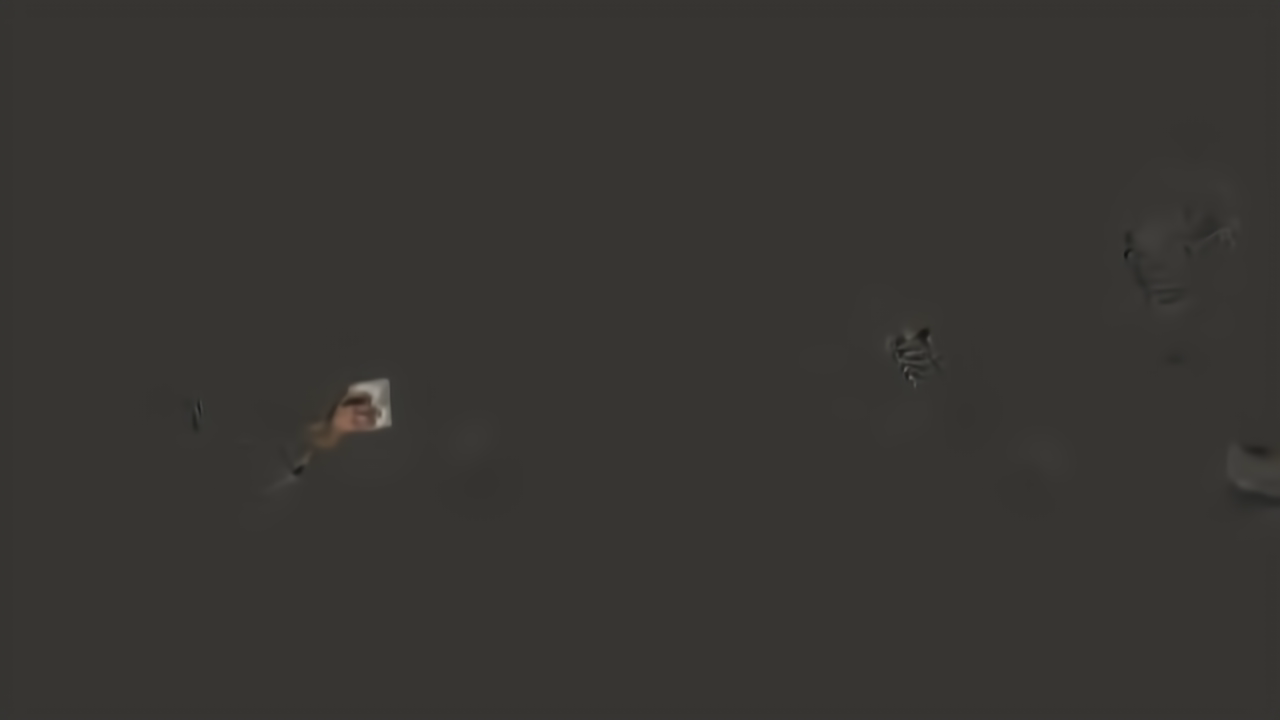}};
            \spy[size=3.5cm, connect spies] on (-1.8,-0.35) in node [left] at (-3.1,0);
            \spy[size=3.5cm, connect spies] on (1.8,0) in node [left] at (6.5,0);
        \end{tikzpicture}
        \caption{CodecNet output from transmitted latents only: $g_s(\hat{\mathbf{y}}_c; \boldsymbol{\theta}_c)$}
        \label{fig:4pcodecsent}  
    \end{subfigure}
    \caption{Illustration of the conditional coding behavior for CodecNet.}
\end{figure}

Conditional coding relevance is illustrated for CodecNet by synthesizing its
output from the shortcut latents only \ref{fig:4pcodecshortcut}, the sent
latents only \ref{fig:4pcodecsent} or both latents \ref{fig:4pcodecall}.
Similarly to MOFNet, most of CodecNet output is retrieved from the shortcut
latents and few information are transmitted, resulting in significant rate
savings. The shortcut latents are computed from the temporal prediction $\pred$.
Therefore, a poor prediction results in shortcut latents lacking some
information, requiring CodecNet to convey something in these areas. Here they
correspond to the quickly moving objects such as the people's hands, whose
prediction results from badly estimated flows (Fig. \ref{fig:4ppastflow},
\ref{fig:4pfutureflow}). This example shows that even with an inaccurate
$\pred$, CodecNet exploits all information from $\pred$ and only transmits
correction terms to obtain a proper reconstruction.

\newcommand{\shortcutroot}{\imagepath/}

\newpage
\subsection{Visual Comparison}

\begin{figure}[h]
    \centering
    \begin{subfigure}{\textwidth}
        \centering
        \begin{tikzpicture}[spy using outlines={circle,palet_red,magnification=5,size=3cm, ultra thick}]
            \node {\includegraphics[width=0.6\linewidth]{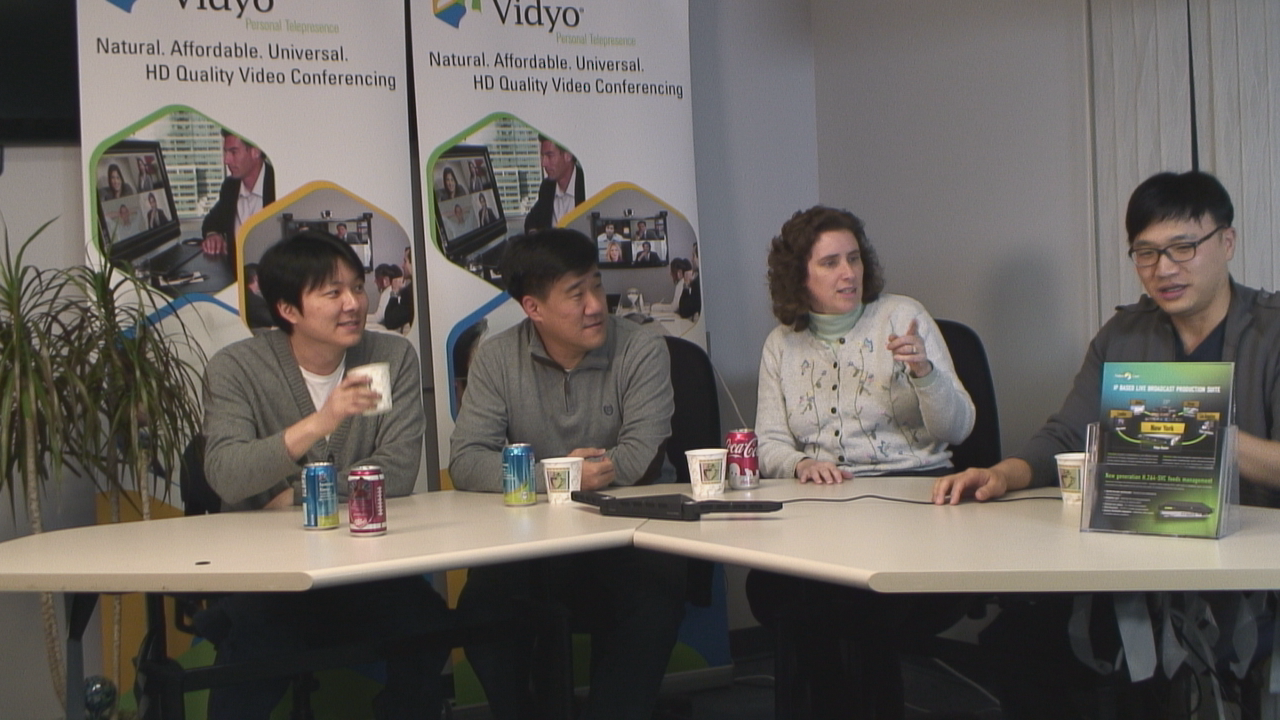}};
            \spy[size=4cm, connect spies] on (-1.9,-0.2) in node [left] at (-3.1,0);
            \spy[size=4cm, magnification=6, connect spies] on (-0.4,1.95) in node [left] at (6,0);
        \end{tikzpicture}
        \caption{Original frame}
      \end{subfigure}

      \medskip
      \begin{subfigure}{\textwidth}
        \centering
        \begin{tikzpicture}[spy using outlines={circle,palet_red,magnification=5,size=3cm, connect spies, ultra thick}]
            \node {\includegraphics[width=0.6\linewidth]{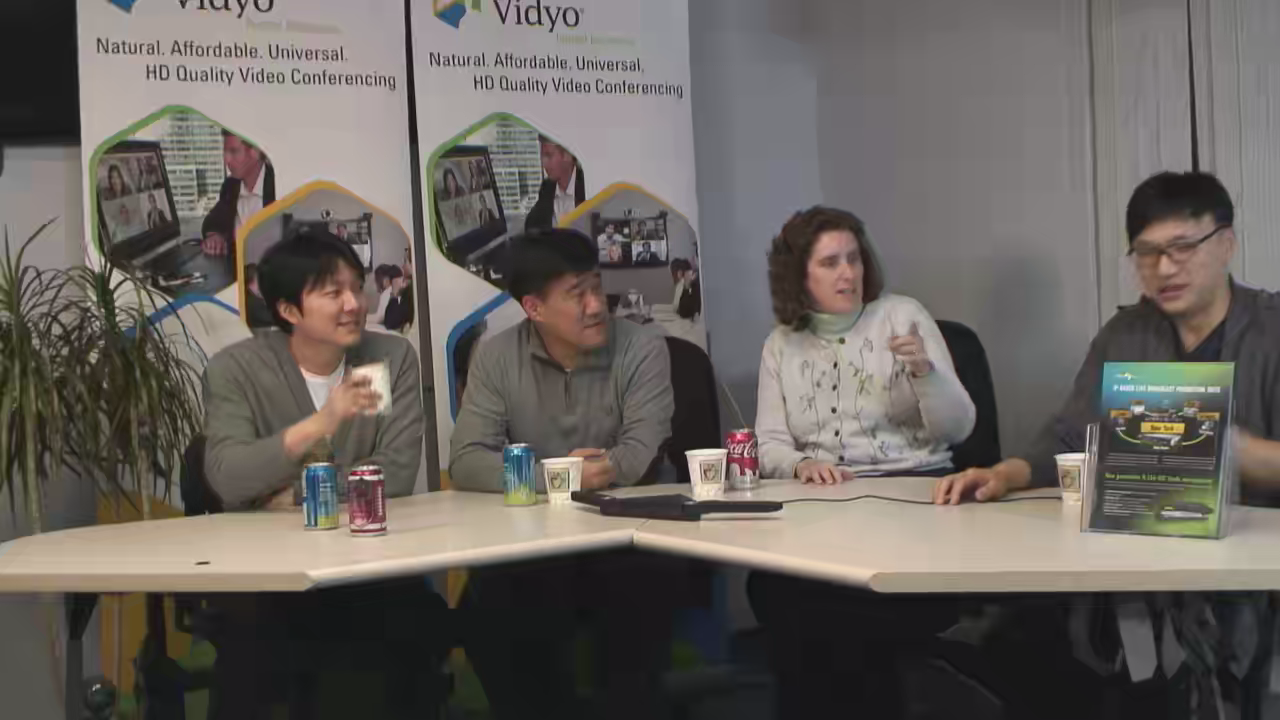}};
            \spy[size=4cm, connect spies] on (-1.9,-0.2) in node [left] at (-3.1,0);
            \spy[size=4cm, magnification=6, connect spies] on (-0.4,1.95) in node [left] at (6,0);
        \end{tikzpicture}
        \caption{HEVC: PSNR of the GOP is $36.94 \text{ dB}$ and the GOP rate is $1.27 \text{ Mbit/s}$. }
      \end{subfigure}

      \medskip
      \begin{subfigure}{\textwidth}
        \centering
        \begin{tikzpicture}[spy using outlines={circle,palet_red,magnification=5,size=3cm, connect spies, ultra thick}]
            \node {\includegraphics[width=0.6\linewidth]{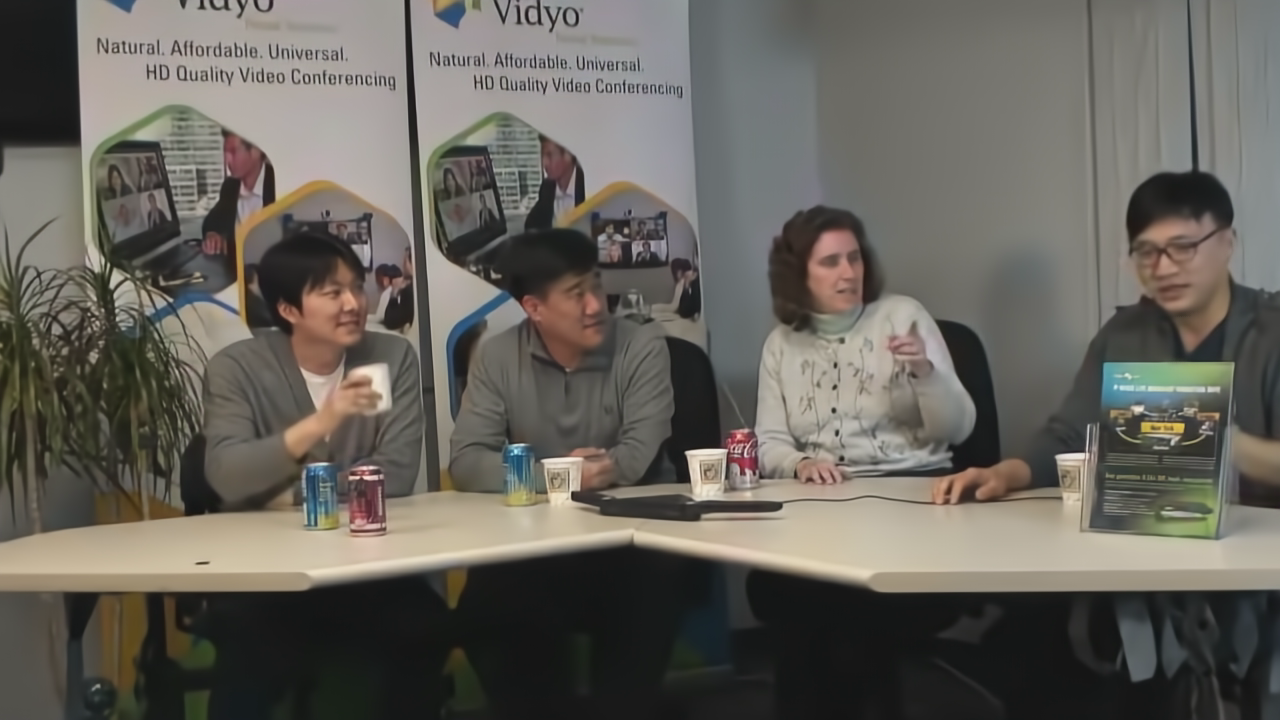}};
            \spy[size=4cm, connect spies] on (-1.9,-0.2) in node [left] at (-3.1,0);
            \spy[size=4cm, magnification=6, connect spies] on (-0.4,1.95) in node [left] at (6,0);
        \end{tikzpicture}
        \caption{Proposed system: PSNR of the GOP is $37.29 \text{ dB}$ and the GOP rate is $1.21 \text{ Mbit/s}$.}
      \end{subfigure}
    \caption{Visual comparison of a B-frame compression.}
    \label{fig:visualcomparison4p}
\end{figure}

The Figure \ref{fig:visualcomparison4p} offers a visual comparison of a B-frame,
compressed by HEVC \footnote{$QP = 35$} and by the proposed system. For a lower rate, the system
achieves a higher PSNR than HEVC. The zoom on the man's hand shows that moving areas
are well handled by the system. The high frequencies in the background
(the text) are properly recovered. The system
obtains a smoother reconstruction with fewer coding artifacts than HEVC,
\textit{i.e.} without blocking or rigging effects.

%

\newcommand{\rawframeroot}{\imagepath/}
\newcommand{\hevcroot}{\imagepath/}
\newcommand{\nnroot}{\imagepath/}
\newcommand{\tikzbelowcaption}{1.5cm}

\newpage
\section{System Behavior on the \textit{BQMall} Sequence}

The behavior of the proposed system is detailed on the \textit{BQMall} sequence,
extracted from the HEVC Common Test Conditions. This sequence features people
walking in front of a static background. In this example, the first frame is
coded as an I-frame and the next 8 frames are compressed using a (random access)
GOP of size 8. This appendix provides additional illustrations to the ones
already shown in section \ref{sec:visualization}.

\subsection{Rate-distortion curves}

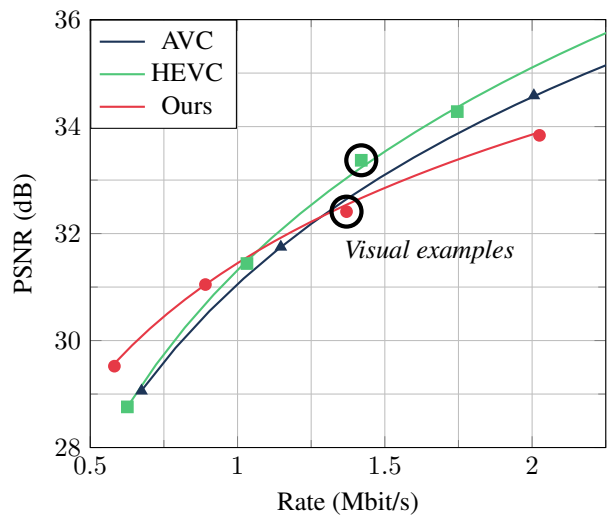
\begin{figure}[h]
    \centering
    \begin{tikzpicture}
        \begin{axis}[
            grid= both ,
            xlabel = {Rate (Mbit/s)} ,
            ylabel = {PSNR (dB)} ,
            ylabel near ticks,
            xlabel near ticks,
            x tick label style={
                /pgf/number format/.cd,
                fixed,
                precision=2
                },
            xtick distance={0.5},
            ytick distance={2},
            minor y tick num=1,
            minor x tick num=1,
            xmin=0.5,xmax=2.25,ymin=28,ymax=36,
            title={Rate-distortion curves on \textit{BQMall} --- BD-rate is -2.6\% w.r.t. HEVC},
            legend style={at={(0.0,1.0)},anchor=north west}
        ]

        \addplot[solid, thick, mark=triangle*, palet_dark_blue, only marks, forget plot] coordinates{
            (0.674,29.06)
            (1.147,31.75)
            (2.006,34.58)
            (3.447,37.29)
            };
        \addplot[thick, domain=0.674:3.447,palet_dark_blue] {5.04484*ln(x) + 31.055869};
        \addlegendentry{AVC};

        \addplot[thick, solid, mark=square*, emerald,only marks, forget plot] coordinates {
            (0.626,28.76)
            (1.032,31.44)
            (1.42,33.37)
            (1.746,34.28)
            (2.973,37.02)
        };
        \addplot[thick, domain=0.626:2.973,emerald] {5.4617*ln(x) + 31.3193};
        \addlegendentry{HEVC};

        \addplot[thick, solid, palet_red, mark=*,only marks, forget plot] coordinates {
            (0.582,29.521)
            (0.891,31.047)
            (1.37,32.41)
            (2.025,33.836)
        };
        \addplot[thick, domain=0.582:2.025,palet_red] {3.46857*ln(x) + 31.447362};
        \addlegendentry{Ours};

        \node (mark) [thick, draw, black, circle, minimum size = 10pt, inner sep=4pt, ultra thick]
            at (axis cs: 1.37,32.41) [] {};
            \node (mark) [thick, draw, black, circle, minimum size = 10pt, inner sep=4pt, ultra thick]
            at (axis cs: 1.42,33.37) [] {};
        
        \node at (axis cs: 1.37,32.41) [below=0.25cm, xshift=1.1cm] {\textit{Visual examples}};

        \end{axis}
    \end{tikzpicture}
    \caption{Rate-distortion curve of the proposed system against AVC and HEVC
    for the \textit{BQMall} sequence. The circled points are
    used to generate the visual examples.}
    \label{fig:bqmallrd}
\end{figure}

The Figure \ref{fig:bqmallrd} presents the rate-distortion results of the
proposed coder against HEVC and AVC. For this sequence the system outperforms
HEVC at low rate. However, the system quality starts saturating at high rate
resulting in worse performance than HEVC. We note that the quality saturation
issue seems to be inherent to the auto-encoder architecture as noted
by \citet{helminger2020lossy}.

\newpage
\subsection{Rate Distribution inside a GOP}

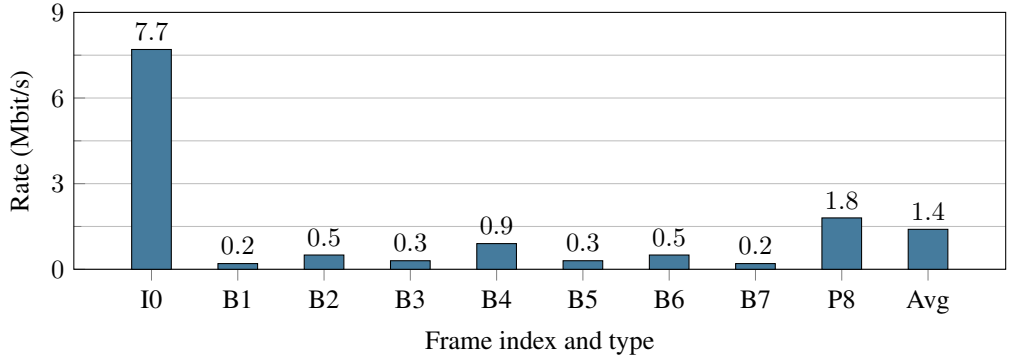
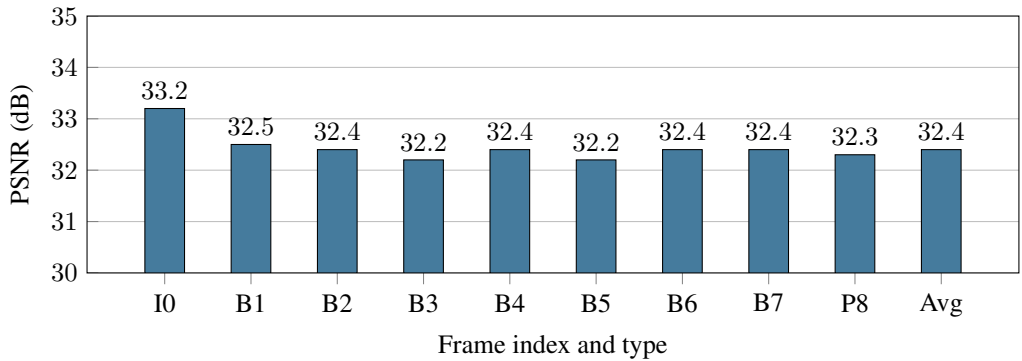
\begin{figure}[H]
    \centering
    \begin{subfigure}{\textwidth}
        \pgfplotsset{compat=1.6}
        \begin{tikzpicture}
            \begin{axis}[
                height=5cm,
                width=\linewidth,
                ylabel=Rate (Mbit/s),
                xlabel=Frame index and type,
                xmin=0, xmax=9.00,
                ymin=0, ymax=9.00,
                ytick distance=3,
                yminorgrids=true,
                ymajorgrids=true,
                minor y tick num=1,
                ybar, bar width=15,
                enlarge x limits=0.1,
                xticklabels={I0,B1,B2,B3,B4,B5,B6,B7,P8, Avg},
                xtick={0,...,9},
                xtick pos=left,
                ytick pos=left,
                nodes near coords,
                nodes near coords align={vertical},
            ]
            \addplot[ybar, fill=palet_medium_blue] coordinates {
                    (0,7.7)
                    (1,0.2)
                    (2,0.5)
                    (3,0.3)
                    (4,0.9)
                    (5,0.3)
                    (6,0.5)
                    (7,0.2)
                    (8,1.8)
                    (9,1.4)
                };
            \end{axis}
        \end{tikzpicture}
    \caption{Rate per frame in a GOP.}
    \end{subfigure}\hfil 

    \medskip
    \begin{subfigure}{\textwidth}
        \pgfplotsset{compat=1.6}
        \begin{tikzpicture}
            \begin{axis}[
                height=5cm,
                width=\linewidth,
                ylabel=PSNR (dB),
                xlabel=Frame index and type,
                xmin=0, xmax=9.00,
                ymin=30, ymax=35,
                ytick distance=1,
                yminorgrids=true,
                ymajorgrids=true,
                minor y tick num=0,
                ybar, bar width=15,
                enlarge x limits=0.1,
                xticklabels={I0,B1,B2,B3,B4,B5,B6,B7,P8, Avg},
                xtick={0,...,9},
                xtick pos=left,
                ytick pos=left,
                nodes near coords,
                nodes near coords align={vertical},
            ]
            \addplot[ybar, fill=palet_medium_blue] coordinates {
                    (0,33.2)
                    (1,32.5)
                    (2,32.4)
                    (3,32.2)
                    (4,32.4)
                    (5,32.2)
                    (6,32.4)
                    (7,32.4)
                    (8,32.3)
                    (9,32.4)
                };
            \end{axis}
        \end{tikzpicture}
    \caption{PSNR per frame in a GOP.}
    \end{subfigure}
    \caption{Distribution of the rate and the PSNR across all frames of a GOP.
    \textit{Avg} denotes the mean value computed on the I-frame and the GOP of size 8.}
    \label{fig:gopdetailbq}
\end{figure}

The Figure \ref{fig:gopdetailbq} presents the distribution of the rate and of
the PSNR across the frames of a GOP. Similarly to the \textit{FourPeople}
sequence, the PSNR remains consistent for all the coded frames. Because this
sequence is less static than \textit{FourPeople}, the inter-frame rates are higher.

\newpage
\subsection{Visual Comparison}

\begin{figure}[H]
    \centering
    \begin{subfigure}{\textwidth}
        \centering
        \begin{tikzpicture}[spy using outlines={circle,palet_red,magnification=5,size=3cm, ultra thick}]
            \node {\includegraphics[width=0.6\linewidth]{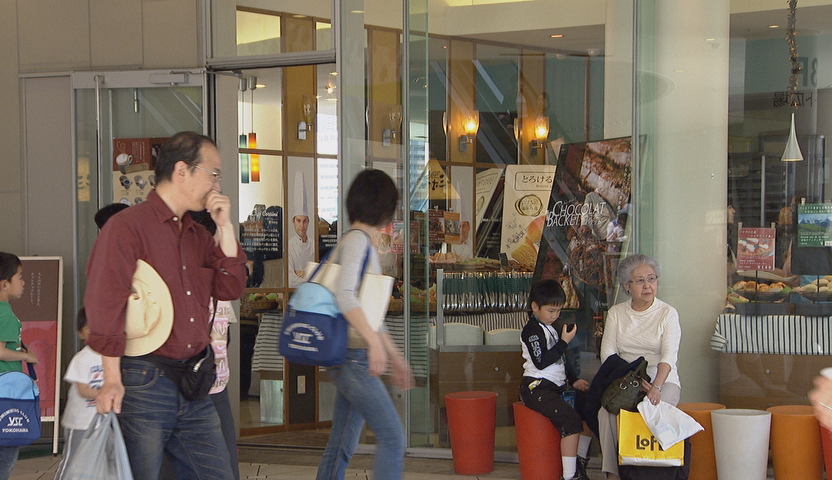}};
            \spy[size=4cm, connect spies] on (-2.1,0.4) in node [left] at (-2.6,0);
            \spy[size=4cm, connect spies] on (-0.55,0.4) in node [left] at (6,0);
        \end{tikzpicture}
        \caption{Original frame}
      \end{subfigure}

      \medskip
      \begin{subfigure}{\textwidth}
        \centering
        \begin{tikzpicture}[spy using outlines={circle,palet_red,magnification=5,size=3cm, connect spies, ultra thick}]
            \node {\includegraphics[width=0.6\linewidth]{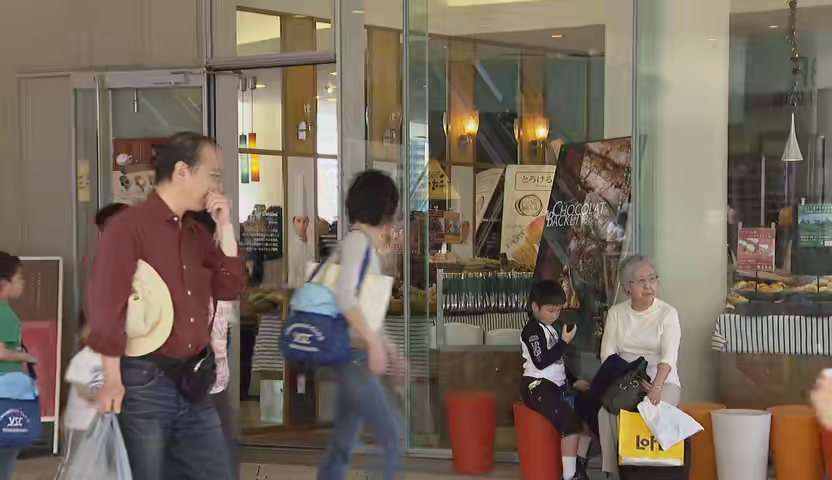}};
            \spy[size=4cm, connect spies] on (-2.1,0.4) in node [left] at (-2.6,0);
            \spy[size=4cm, connect spies] on (-0.55,0.4) in node [left] at (6,0);
        \end{tikzpicture}
        \caption{HEVC: The PSNR of the GOP is $33.37 \text{ dB}$ and the GOP rate is $1.42 \text{ Mbit/s}$. }
      \end{subfigure}

      \medskip
      \begin{subfigure}{\textwidth}
        \centering
        \begin{tikzpicture}[spy using outlines={circle,palet_red,magnification=5,size=3cm, connect spies, ultra thick}]
            \node {\includegraphics[width=0.6\linewidth]{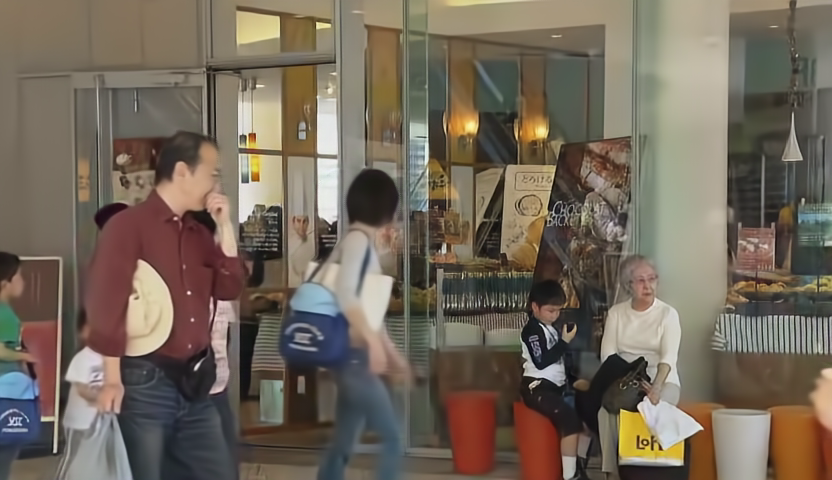}};
            \spy[size=4cm, connect spies] on (-2.1,0.4) in node [left] at (-2.6,0);
            \spy[size=4cm, connect spies] on (-0.55,0.4) in node [left] at (6,0);
        \end{tikzpicture}
        \caption{Proposed system: The PSNR of the GOP is $32.41 \text{ dB}$ and the GOP rate is $1.37 \text{ Mbit/s}$.}
      \end{subfigure}
      \medskip
    \caption{Visual comparison of a B-frame compression.}
    \label{fig:visualcomparisonbq}
\end{figure}

The Figure \ref{fig:visualcomparisonbq} offers a visual comparison of a B-frame,
compressed by HEVC\footnote{$QP = 34$} and by the proposed system. At a similar rate, HEVC achieves
a better PSNR than the proposed coder and seems to retain more high frequency
contents. However, it comes as the cost of significant blocking artifacts and
pronounced ringing effects. Due to its convolutional nature, the proposed system
offers a smoother output, with fewer compression artifacts.

\renewcommand{\rawframeroot}{\imagepath/}
\renewcommand{\hevcroot}{\imagepath/}
\renewcommand{\nnroot}{\imagepath/}


\newpage
\section{Description of the Network Architecture}
\label{app:detailarchitecture}

The architecture of MOFNet and CodecNet, presented in Figure
\ref{fig:overalldiagram}, are described in this appendix.

\subsection{Basic Building Blocks}

\begin{figure}[H]
    \centering 
    \begin{subfigure}[b]{0.4\textwidth}
        \includegraphics[width=\linewidth]{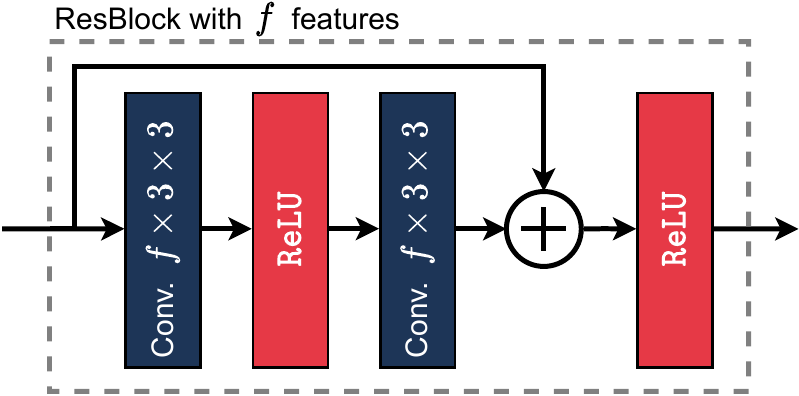}
        \caption{Residual block}
    \end{subfigure}\hfil 
    \begin{subfigure}[b]{0.5\textwidth}
        \includegraphics[width=\linewidth]{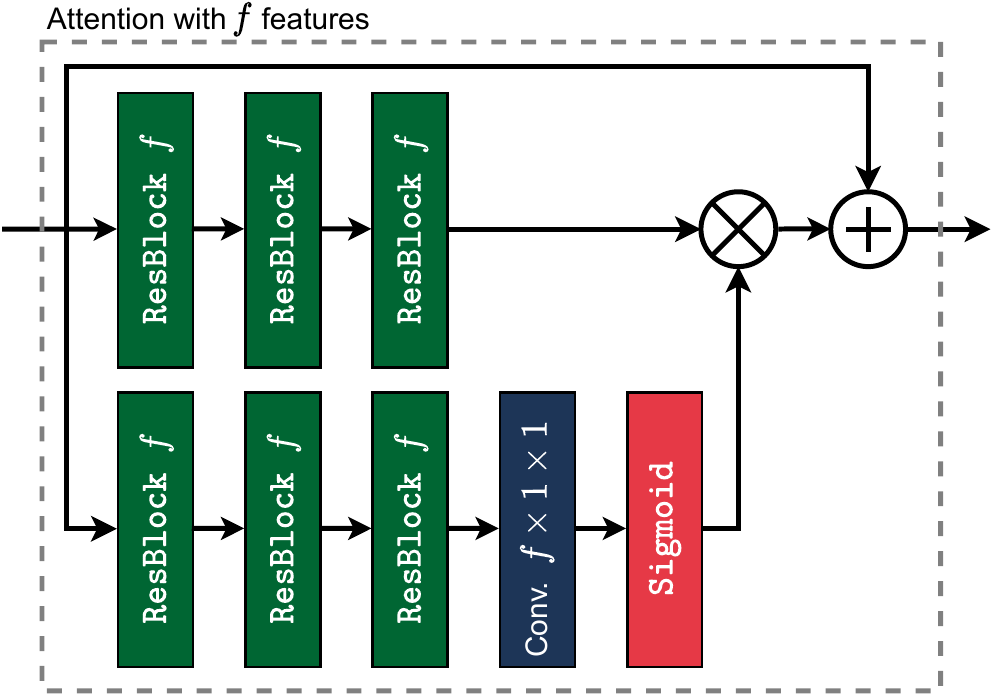}
        \caption{Attention module}
    \end{subfigure}
    \caption{Architecture of an attention module. Conv. $f \times k
    \times k$ denotes a convolutional layer with $f$ output features and a $k \times k$ kernel.}  
    \label{fig:appbuildingblocks}
\end{figure}

The system uses attention module to increase the capacity of its
different transforms. The attention modules are implemented as proposed by
\citet{cheng2020learned} and are described in Figure \ref{fig:appbuildingblocks}.

\newcommand{\detailedarchi}{\imagepath/}

\subsection{Hyperprior and Entropy Coding}

\begin{figure}[H]
    \centering 
    \begin{subfigure}[b]{0.5\textwidth}
        \includegraphics[width=\linewidth]{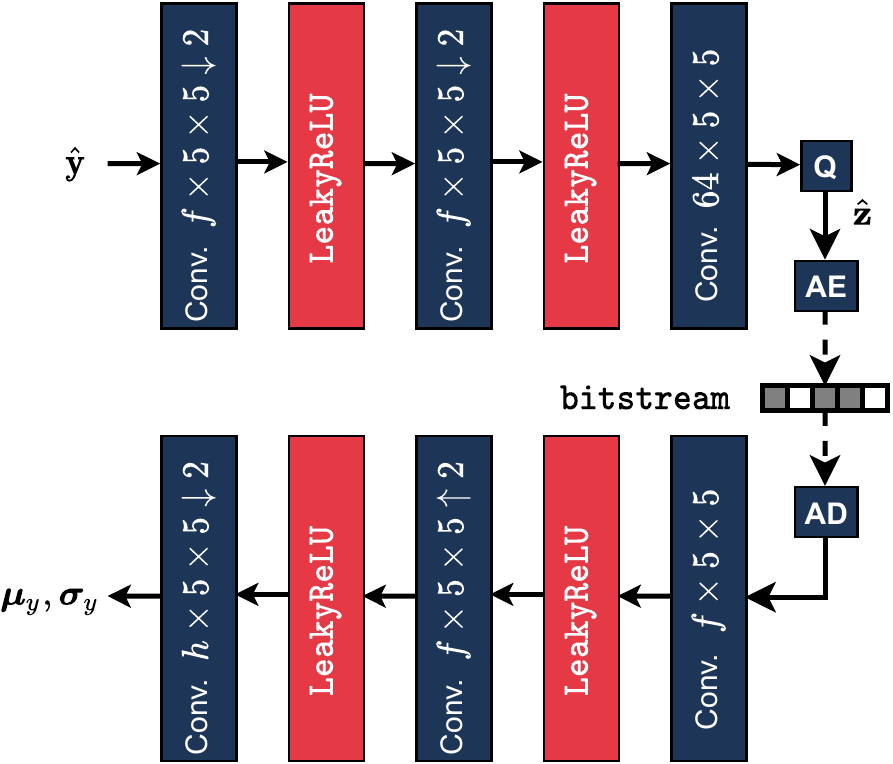}
        \caption{Hyperprior autoencoder}
        \label{fig:apphyperprior}
    \end{subfigure}\hfil 
    \begin{subfigure}[b]{0.35\textwidth}
        \includegraphics[width=\linewidth]{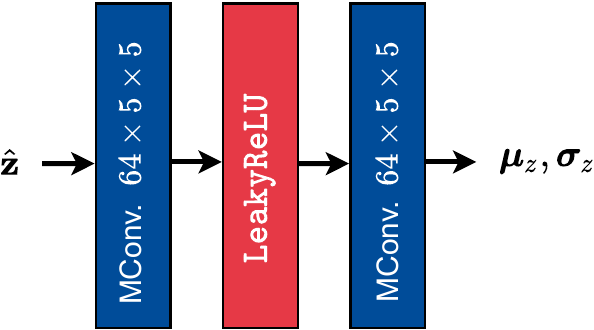}
        \caption{Auto-regressive model.}
        \label{fig:arm}
    \end{subfigure}
    \caption{Architecture of the hyperprior network. $\hat{\mathbf{y}}$
    corresponds to MOFNet (sent) latents $\hat{\mathbf{y}}_m$
    or to CodecNet (sent) latents $\hat{\mathbf{y}}_c$. Conv. $f
    \times k \times k\ \uparrow/\downarrow 2$ denotes a convolutional layer with
    $f$ output features, a $k \times k$ kernel and a up/down sampling by a factor
    2. MConv is a masked convolution.}  
\end{figure}

The transmitted latents of MOFNet and CodecNet are conveyed using entropy
coding, which requires an estimate of the latents probability density function
(PDF). Each element $\hat{y}_i$ of the latents is described by a Laplace PDF,
whose parameters $\mu_i, \sigma_i$ are conditioned on a hyperprior
$\hat{\mathbf{z}}$ \citep{DBLP:conf/iclr/BalleMSHJ18}. The hyperprior is
computed and transmitted from an auxiliary auto-encoder, described in Figure
\ref{fig:apphyperprior}. The hyperprior transmission uses entropy coding
and a Laplace PDF, whose parameters are estimated with an auto-regressive
model (see Fig. \ref{fig:arm}) as proposed by \citet{DBLP:conf/nips/MinnenBT18}.
Two hyperprior networks are implemented, one for MOFNet and one for CodecNet.

\newpage
\subsection{MOFNet Architecture}

\begin{figure}[H]
    \begin{subfigure}{0.8\textwidth}
        \includegraphics[width=\linewidth]{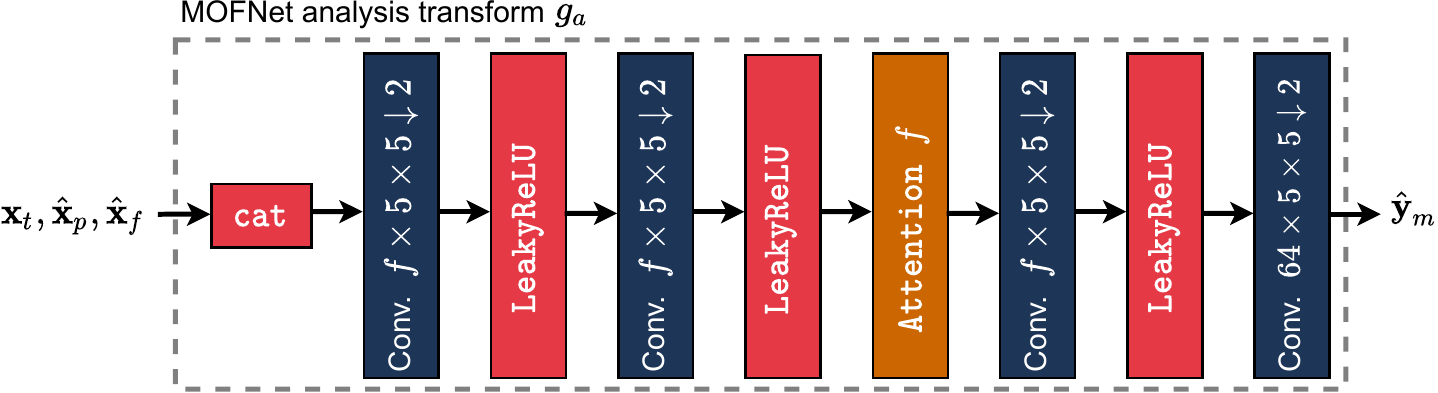}
        \caption{MOFNet analysis transform $g_a$}
    \end{subfigure}

    \medskip
    \begin{subfigure}{0.8\textwidth}
        \includegraphics[width=\linewidth]{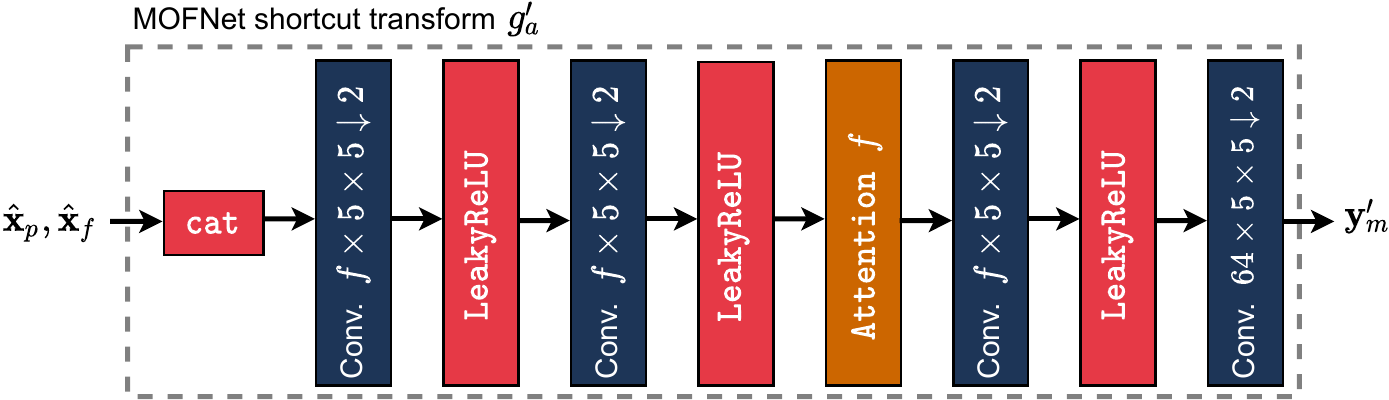}
        \caption{MOFNet shortcut transform $g_a^\prime$}
    \end{subfigure}

    \medskip
    \begin{subfigure}{\textwidth}
        \includegraphics[width=\linewidth]{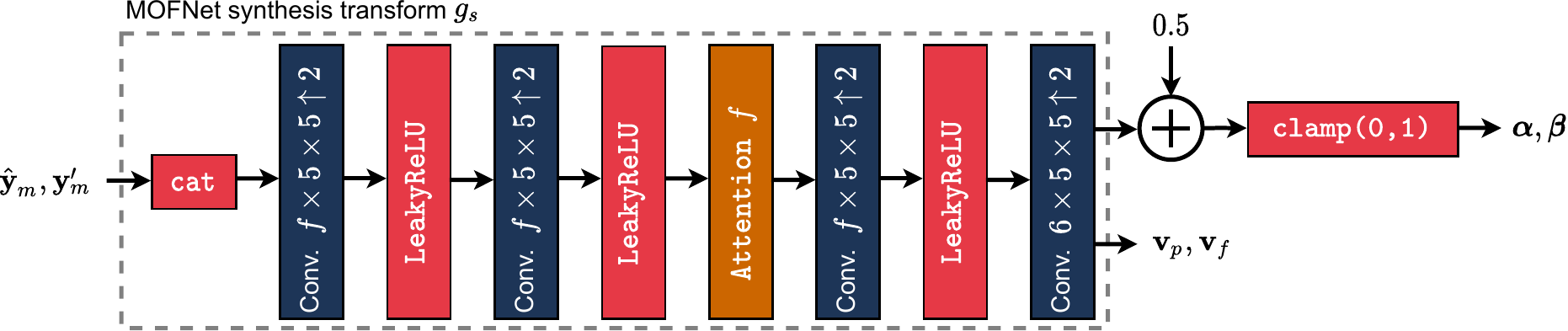}
        \caption{MOFNet synthesis transform $g_s$}
    \end{subfigure}
    \caption{MOFNet transforms architecture. Conv. $f \times k \times k\
    \uparrow/\downarrow 2$ denotes a convolutional layer with $f$ output
    features, a $k \times k$ kernel and a up/down sampling by a factor 2.
    $\texttt{Attention}\ f$ is an attention module with $f$ features,
    $\texttt{cat}$ represents the concatenation along the features dimension and
    $\texttt{clamp}(0,1)$ is a hard clipping between $0$ and $1$. $f$ is set to
    128.}
    \label{fig:appmofnetdetail}
\end{figure}

The detailed architecture of the three main transforms of MOFNet (analysis,
shortcut and synthesis) is depicted in Figure \ref{fig:appmofnetdetail}.

\newpage
\subsection{CodecNet Architecture}

\begin{figure}[H]
    \begin{subfigure}{0.8\textwidth}
        \includegraphics[width=\linewidth]{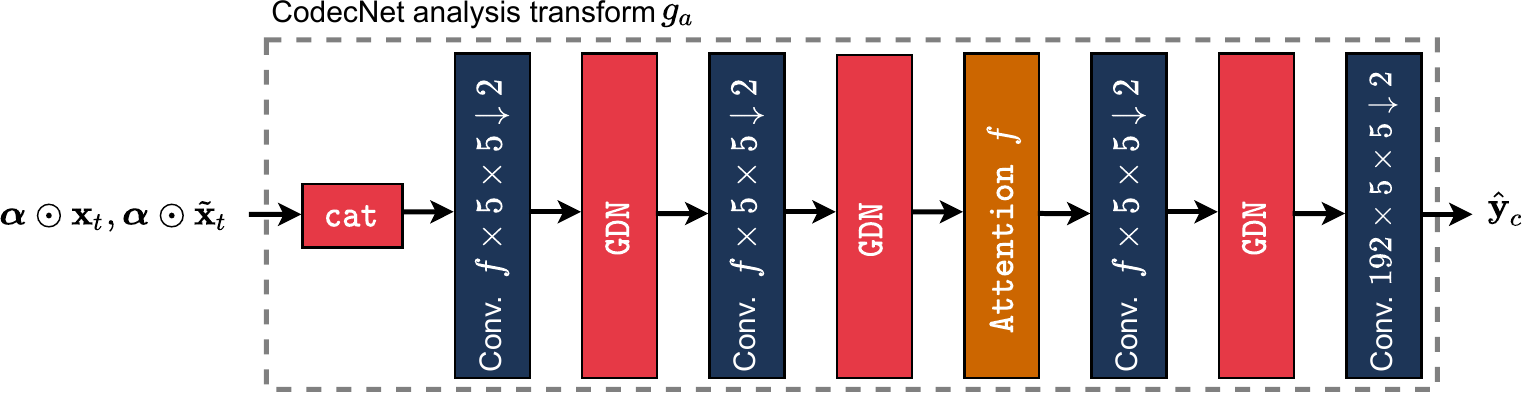}
        \caption{CodecNet analysis transform $g_a$}
    \end{subfigure}

    \medskip
    \begin{subfigure}{0.8\textwidth}
        \includegraphics[width=\linewidth]{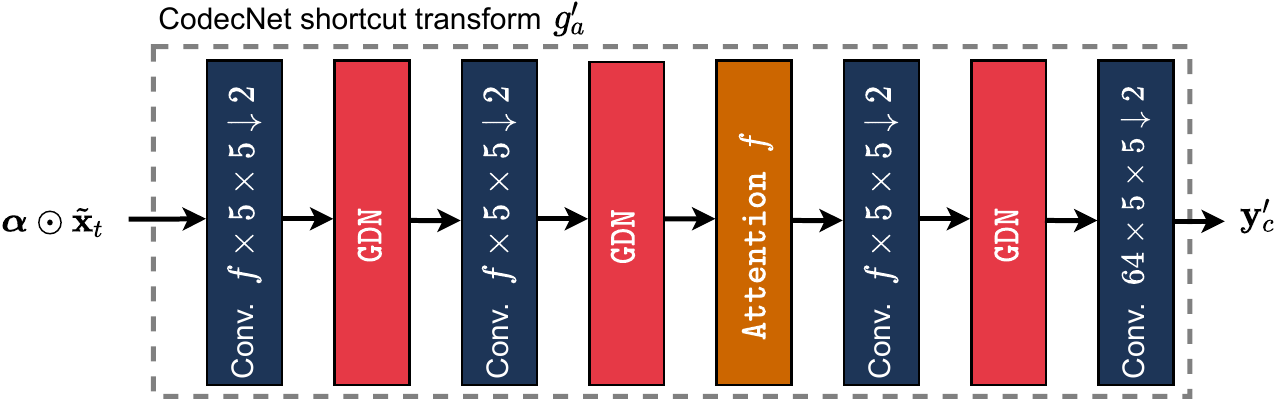}
        \caption{CodecNet shortcut transform $g_a^\prime$}
    \end{subfigure}

    \medskip
    \begin{subfigure}{0.8\textwidth}
        \includegraphics[width=\linewidth]{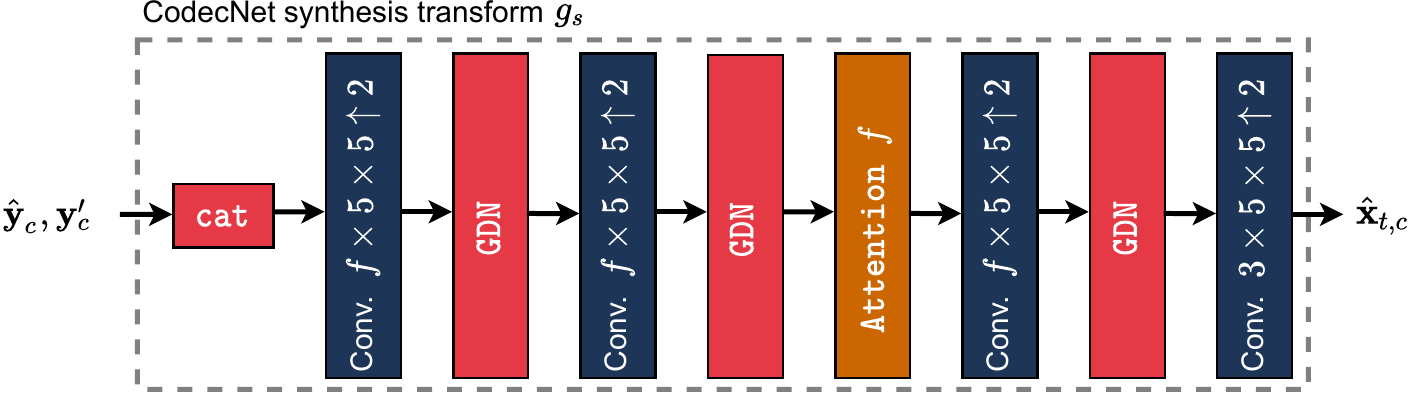}
        \caption{CodecNet synthesis transform $g_s$}
    \end{subfigure}
    \caption{CodecNet transforms architecture. Conv. $f \times k \times k\
    \uparrow/\downarrow 2$ denotes a convolutional layer with $f$ output
    features, a $k \times k$ kernel and a up/down sampling by a factor 2.
    $\texttt{Attention}\ f$ is an attention module with $f$ features,
    $\texttt{cat}$ represents the concatenation along the features dimension.
    $\texttt{GDN}$ is the General Divisive Normalization introduced by
    \citep{DBLP:conf/iclr/BalleLS17}. $f$ is set to 128.}
    \label{fig:appcodecnetdetail}

\end{figure}

The detailed architecture of the three main transforms of CodecNet (analysis, shortcut and synthesis) is depicted in Figure \ref{fig:appcodecnetdetail}.

\end{document}